\documentclass{jpp}
\usepackage{hyperref}
\usepackage{url}
\usepackage{graphicx}
\usepackage{subcaption}
\usepackage[utf8]{inputenc}
\usepackage[T1]{fontenc}
\usepackage{amsmath}
\usepackage{breqn}
\newcommand{\partder}[2]{\frac{\partial #1}{\partial #2}}

\usepackage{bm}
\usepackage{xcolor}

\newcommand{\deltan}{\delta\!n}
\newcommand{\deltam}{\delta\!m}
\newcommand{\deltambar}{\delta\!\overline{m}}
\newcommand{\deltashort}{\delta\!}

\newcommand{\Deltam}{\Delta m}
\newcommand{\Deltan}{\Delta n}
\newcommand{\omegazero}{\overline{\omega}^{(0)}}

\shorttitle{}
\shortauthor{}

\title{The shear Alfv\'{e}n continuum of quasisymmetric stellarators}
\shorttitle{Shear Alfv\'{e}n continuum of QS stellarators}

\author{Elizabeth J. Paul\aff{1}\corresp{\email{ejp2170@columbia.edu}}, Abdullah Hyder\aff{1}, Eduardo Rodr\'{i}guez\aff{2}, Rog\'{e}rio Jorge\aff{3}, Alexey Knyazev\aff{1}}
\shortauthor{E. J. Paul and coauthors}

\affiliation{\aff{1}Department of Applied Physics and Applied Mathematics, Columbia University, New York, NY 10027, USA \aff{2} Max Planck Institute for Plasma Physics, 17491 Greifswald, Germany \aff{3} Department of Physics, University of Wisconsin-Madison, Madison, WI 53706, USA}

\begin{document}

\maketitle

\begin{abstract}
The shear Alfvén wave (SAW) continuum plays a critical role in the stability of energetic particle-driven Alfv\'{e}n eigenmodes. We develop a theoretical framework to analyze the SAW continuum in three-dimensional quasisymmetric magnetic fields, focusing on its implications for stellarator design. By employing a near-axis model and degenerate perturbation theory, the continuum equation is solved, highlighting unique features in 3D configurations, such as the interactions between spectral gaps. Numerical examples validate the theory, demonstrating the impact of flux surface shaping and quasisymmetric field properties on continuum structure. The results provide insights into optimizing stellarator configurations to minimize resonance-driven losses of energetic particles. This work establishes a basis for incorporating Alfvénic stability considerations into the stellarator design process, demonstrated through optimization of a quasihelical configuration to avoid high-frequency spectral gaps. 
\end{abstract}

\section{Introduction}

Shear Alfv\'{e}n waves (SAW) are low-frequency, incompressible, MHD oscillations associated with field-line bending \citep{1995Chen}. 
Given typical values of the phase velocity $v_A = B/\sqrt{\mu_0 \rho}$, where $B$ is the magnetic field strength and $\rho$ is the density, SAWs have the potential for resonant interactions with energetic particle (EP) populations in fusion plasmas \citep{1995Chen}. 
In an inhomogeneous plasma, a so-called continuous spectrum of frequencies exists, consisting of radially singular solutions in the SAW eigenvalue equation. 
Wave packets excited in the continuum are often strongly damped due to phase mixing; therefore, the most easily excited modes reside in the continuum frequency gaps. For this reason, the calculation of the SAW continuum is often the first step in predicting the discrete frequency spectrum and stability of EP-driven Alfv\'{e}n eigenmodes. These gaps exist because counter-propagating SAW waves are coupled through the poloidal and toroidal variation of the magnetic geometry, analogous to the existence of electron band gaps due to periodic modulations of the potential \citep{2008Heidbrink}. 

In axisymmetric geometry, the toroidal  Alfv\'{e}n eigenmode (TAE) and the ellipticity-induced  Alfv\'{e}n eigenmode (EAE) exist because of $m = 1$ and $m =2$ dependence of the magnetic geometry, respectively, where $m$ is the poloidal mode number. These modes are widely observed in both tokamak and stellarator experiments \citep{2006Van,2011Toi,2014Gorelenkov}. 
In 3D systems, additional gaps arise due to the dependence of the geometry on the toroidal angle, giving rise to the helical Alfv\'{e}n eigenmode (HAE)---corresponding to $m \ne 0$ and $n \ne 0$---and the mirror Alfv\'{e}n eigenmode (MAE)---corresponding to $n \ne 0$ and $m = 0$ \citep{2003Spong,2001Kolesnichenko}, where $n$ is the toroidal mode number. 

Stellarators can be designed to be quasisymmetric, with a Noether symmetry of the guiding center Lagrangian implying excellent neoclassical confinement \citep{1988Nuhrenberg,1983Boozer,2022Landreman}. However, the flux-surface shaping does not inherit the same symmetry as the field strength. Therefore, even if the level of quasisymmetry is very precise, the gap structure of a quasisymmetric stellarator can differ substantially from that of a tokamak. Furthermore, 3D geometry yields distinct features of continuum eigenfunctions, leading to modes that can become localized within the magnetic surface \citep{2007Yakovenko} or the formation of gaps that overlap and interact with one another \citep{2011Kolesnichenko}. 

Energetic particles have historically been challenging to confine in stellarator configurations due to the possibility of unconfined orbits and resonances exposed at low collisionality \citep{1992Grieger,1999Fedi}. These difficulties must be overcome to develop a stellarator reactor concept, as excessive alpha losses before thermalization can impact power balance and impart damage to plasma-facing components. When 3D fields are introduced, such as in a stellarator or rippled tokamak, the collisionless guiding center orbits are no longer automatically well confined. This leads to increased collisionless losses---due to ripple trapping, collisionless diffusion and drift-convective orbits \citep{2001Beidlerb,2006Mynick,2022Paul}---and weakly collisional transport is generally enhanced. If a stellarator magnetic field is sufficiently close to quasisymmetry, the conservation of the corresponding canonical momentum \citep{2020Rodriguez} provides guiding center confinement if the orbit width is sufficiently small. Recent optimization studies have revealed stellarator configurations with precise levels of quasisymmetry, yielding confinement of alpha particle trajectories of similar levels to that in tokamaks \citep{2022Landreman,2022Landremanb,2024Nies}. 
With the possibility that guiding center orbits can be confined, there is, however, the potential for enhanced alpha losses due to interactions with SAWs. While interaction between EPs and other MHD modes, e.g. kink and sawtooth, is possible, Alfv\'{e}nic activity is considered the major limitation to alpha confinement in burning toroidal plasma \citep{2014Gorelenkov}. 

While the SAW continuum has been calculated numerically for several quasisymmetric equilibria \citep{2003Spong,2004Fesenyuk,2021Varela}, the goal of this work is to study the underlying features of the shear Alfv\'{e}n continuum in quasisymmetric configurations. 
With an improved understanding of the continuum, we hope to better assess the potential impact of AE instabilities in quasisymmetric devices and account for AE stability in the stellarator design process. 

In Section \ref{sec:saw_overview} we outline the SAW model in 3D magnetic fields. A near axis model magnetic field \citep{1991Garren,2019Landremanc}, discussed in Section \ref{sec:nae_qs}, is applied to elucidate the impact of quasisymmetry on the flux-surface shaping, and therefore the continuum structure. Given the common physics basis between the shear Alfv\'{e}n continuum and the formation of band gaps, degenerate perturbation theory is used to solve the continuum equation in Section \ref{sec:perturbative_solution}. We highlight key features of stellarator continuum solutions, such as higher-order crossings and the possibility of co-propagating modes which appear to cross spectral gaps. Numerical examples are shown in Section \ref{sec:numerical_continuum} to validate the theory. In Section \ref{sec:gap_resonance}, we discuss the impact of the SAW gap location on the passing EP resonance condition. Finally, in Section \ref{sec:optimization} we demonstrate an optimization strategy to manipulate the SAW continuum to avoid passing EP resonances.

\section{Shear Alfv\'{e}n continuum in 3D geometry}
\label{sec:saw_overview}

 An equation describing shear Alfv\'{e}n waves is obtained under the assumption of $\beta \ll 1$ and the reduced MHD ordering $\epsilon = k_{\|}/k_{\perp} \sim 1/(k_{\perp} a) \ll 1$ \citep{salat_shear_2001,Fesenyuk2002}. Here $k_{\|} \sim \nabla_{\|} = \hat{\bm{b}} \cdot \nabla $ and $k_{\perp} \sim \nabla_{\perp} = \nabla - \hat{\bm{b}} \nabla_{\|}$ represent typical wave numbers of perturbed quantities parallel and perpendicular to the equilibrium magnetic field, and $\beta = p/(B^2/2\mu_0)$ is the ratio of the thermal pressure to the magnetic pressure. 
 The linearized quasineutrality condition yields a PDE for the perturbed electrostatic potential $\Phi$,
\begin{align}
       B \nabla_{\|} \left(\frac{  \nabla_{\perp}^2 \left(\nabla_{\|} \Phi\right)}{B} \right) + \frac{\omega^2}{v_A^2} \nabla_{\perp}^2 \Phi = 0,
       \label{eq:SAW}
\end{align}
where $v_A^2 = B^2/(\mu_0 \rho)$ is the Alfv\'{e}n velocity, $\rho$ is the mass density, and $\omega$ is the frequency. Note that we have neglected the coupling to sound waves, which has been shown to modify low-frequency gaps in stellarators \citep{2010Konies}. This is justified, since we will focus on high-frequency gaps for the following analysis. 

The shear Alfv\'{e}n equation has a continuous spectrum, corresponding to a set of frequencies for which the corresponding eigenfunctions are radially singular. The continuum solutions correspond to a localization of the eigenfunction on a corresponding singular surface. The shear Alfvén continuum equation is obtained from the terms in \eqref{eq:SAW} with the highest-order radial derivative \citep{salat_shear_2001-1}, 
\begin{align}
     B \nabla_{\|} \left(\frac{|\nabla \psi|^2}{B} \nabla_{\|} \Phi\right) + \frac{\omega^2|\nabla \psi|^2}{v_A^2} \Phi = 0.
     \label{eq:SAW_continuum}
\end{align}
Solutions to \eqref{eq:SAW_continuum} can be interpreted as solutions to \eqref{eq:SAW} with delta-function-like radial variation. To analyze the formation of continuum gaps, we write the continuum equation in Boozer coordinates \citep{1981Boozer} $(\psi,\theta,\zeta)$, where $2\pi \psi$ is the toroidal flux and $\theta$ and $\zeta$ are the poloidal and toroidal angle, respectively, and adopt appropriate normalizations:
\begin{align}
    \left(\partder{}{\zeta} + \iota \partder{}{\theta} \right) \left[\frac{|\nabla \psi|^2}{\langle |\nabla \psi|^2 \rangle}
    \left(\partder{}{\zeta} + \iota \partder{}{\theta} \right) \Phi\right] + \frac{\omega^2}{\omega_A^2}\frac{|\nabla \psi|^2/B^4}{\langle |\nabla \psi|^2 \rangle/\langle B^4\rangle} \Phi  = 0.
    \label{eq:SAW_continuum_Boozer}
\end{align}
We have defined the effective Alfv\'{e}n frequency as $\omega_A = \langle B\rangle^2/(G+\iota I)\sqrt{\mu_0 \rho})$, where $\langle \dots \rangle = (4\pi^2)^{-1}\int_0^{2\pi} \int_0^{2\pi} d\zeta d \theta \dots$ indicates an angle average of a geometric quantity. Here $G$ and $I$ are the toroidal and poloidal covariant components of the magnetic field, defined through 
\begin{align}
    \bm{B} = G(\psi) \nabla \zeta + I(\psi) \nabla \theta + K(\psi,\theta,\zeta)\nabla \psi. 
\end{align}
The perturbed potential can be expressed as a Fourier series in Boozer angles,
\begin{align}
    \Phi = \sum_{m,n} \Phi_{m,n} e^{i(m\theta - n \zeta + \omega t)}. 
    \label{eq:deltaphi_Fourier}
\end{align}
As seen in \eqref{eq:SAW_continuum_Boozer}, the equilibrium quantities appearing in the continuum equation, $|\nabla \psi|^2$ and $|\nabla \psi|^2/B^4$, provide coupling between mode numbers of the potential.
In a two-dimensional magnetic field, the toroidal harmonics $n$ are decoupled. Similarly, in a 3D field with field-period symmetry (all equilibrium quantities only depend on the toroidal angle through $N_P\zeta$ where $N_P$ is the number of field periods), toroidal coupling is isolated to mode families, the sets of mode numbers $n$ separated by integer multiples of $N_P$. We will focus on the solutions of \eqref{eq:SAW_continuum_Boozer} in a quasisymmetric (QS) field, for which the magnetic field strength only varies on a magnetic surface through the angle $\chi = \theta - N \zeta$, where $N$ is the symmetry helicity ($N = 0$ for quasiaxisymmetry, QA, and $N = \pm N_P$ for quasihelical symmetry, QH). We note that even if the magnetic field strength is quasisymmetric, implying that $B(\psi,\chi)$, the other geometric quantities, such as $|\nabla \psi|^2$, do not necessarily respect the same symmetry. Thus the continuous spectrum in quasisymmetric stellarators will generally differ dramatically from that of axisymmetric devices. This will be shown explicitly in the following section, using a near-axis model field.

\section{Near-axis quasisymmetry model}
\label{sec:nae_qs}

Further analysis of the continuum equation \eqref{eq:SAW_continuum_Boozer} requires a model for the geometric factors appearing, namely $|\nabla \psi|^2$ and $B$. 
We will adopt the near-axis quasisymmetric magnetic field description \citep{2019Landremanc,1991Garren}, an asymptotic expansion in the distance from the magnetic axis $r = \sqrt{2\psi/B_0}$, where $B_0$ is the field strength on the magnetic axis. Although such a description is limited to some region near the axis (not necessarily small), it has been shown to capture the nature of such fields, and is thus highly insightful \citep{2019Landreman, 2022Landreman_mapping, rodriguez2023constructing}. Furthermore, since the birth rate of alpha particles will be peaked in the core, this model is appropriate for quasisymmetric reactors. 
In such a model, the magnetic field strength, covariant components of the magnetic field, and rotational transform read:
\begin{align}
\left \{
\begin{array}{c}
    B = B_0\left(1 - r \overline{\eta} \cos(\chi)\right) + \mathcal{O}(r^2) \\
    G = G_0 + \mathcal{O}(r^2) \\
    I = r^2I_2 + \mathcal{O}(r^4) \\
    \iota = \iota_0 + \mathcal{O}(r^2)
\end{array}
    \right. ,
\end{align}
where $\overline{\eta}$, $G_0$, and $\iota_0$ are constants that define the magnetic field in Boozer coordinates through the above expressions. Furthermore, the geometric factor appearing in the continuum equation \eqref{eq:SAW_continuum_Boozer} reads \citep{2020Jorge}, 
\begin{align}
    |\nabla \psi|^2 = r^2 \Psi_2 + r^3 \Psi_3 + \mathcal{O}(r^4),
    \label{eq:psi_expansion}
\end{align}
with 
\begin{align}
    \Psi_2 &= \frac{B_0^2}{2\overline{\kappa}^2}\left[1+\overline{\kappa}^4(1+\sigma^2) + \cos(2\chi)\left(-1 + \overline{\kappa}^4(1 - \sigma^2) \right) + \sin(2\chi)\left(-2 \sigma \overline{\kappa}^4\right)\right]
    \label{eq:gradpsi2_nae}
\end{align}
and $\Psi_3$ provided in Appendix \ref{app:third_order_nablapsi2}. Here $\overline{\kappa}(\zeta) = \kappa(\zeta)/\overline{\eta}$ is the normalized magnetic axis curvature and $\sigma(\zeta)$ satisfies the ODE,
\begin{align}
        0 &= \sigma' + (\iota_0-N) \left(\frac{1}{\overline{\kappa}^4} + 1 + \sigma^2 \right) +   2 \left(\tau - \frac{I_2}{B_0}\right) \frac{G_0}{B_0 \overline{\kappa}^2},
        \label{eq:sigma_equation}
\end{align}
where $\tau(\zeta)$ is the magnetic axis torsion and the prime denotes $d/d\zeta$. The function $\sigma(\zeta)$ controls the flux-surface ellipticity and rotation, as will be described in more detail below. 

In summary, with the parameters $\overline{\eta}$, $B_0$, $G_0$, and $I_2$ prescribed along with the shape of the magnetic axis (which determines $N$ \citep{2023Rodriguezb}), the ODE \eqref{eq:sigma_equation} with appropriate boundary conditions \citep{2018Landreman} determines the function $\sigma(\zeta)$ as well as the on-axis rotational transform $\iota$. With these quantities, all geometric information needed for solving the near-axis continuum equation is available. Under the near-axis assumptions, the continuum equation to lowest order in $r$ reads:
\begin{align}
    \frac{1}{\Psi_2}\left(\partder{}{\zeta} + \iota_0 \partder{}{\theta} \right) \left[\Psi_2
    \left(\partder{}{\zeta} + \iota_0 \partder{}{\theta} \right) \Phi\right] + \overline{\omega}^2 \Phi  = 0.
    \label{eq:continuum_gradpsi2}
\end{align}
Here, we define the effective Alfv\'{e}n frequency as $\omega_A = B_0^2/(G_0\sqrt{\mu_0 \rho})$ and the normalized frequency $\overline{\omega} = \omega/\omega_A$. (Note that the geometric factor $G_0/B_0$ defines an effective major radius through $R_0^{-1} = |\nabla \zeta| = B_0/G_0$, using the lowest-order covariant expression for the magnetic field.)
We employ a perturbative approach to solving \eqref{eq:continuum_gradpsi2} for the small parameter $\epsilon$ quantifying the geometric coupling,
\begin{align}
\epsilon = \frac{\Psi_2}{\langle \Psi_2\rangle} -1,
    \label{eq:A_nae}
\end{align}
such that the continuum equation reads,
\begin{align}
    \frac{1}{\left(1  + \epsilon\right)}\left(\partder{}{\zeta} + \iota_0 \partder{}{\theta} \right) \left[\left(1  + \epsilon\right)
    \left(\partder{}{\zeta} + \iota_0 \partder{}{\theta} \right) \Phi\right] + \overline{\omega}^2 \Phi  = 0.
    \label{eq:SAW_continuum_nae}
\end{align}

Representing $\Phi$ as a Fourier series \eqref{eq:deltaphi_Fourier}, \eqref{eq:SAW_continuum_nae} implies that coupling between mode numbers $(m,n)$ of the potential will be provided by the variation of $\epsilon$ on a magnetic surface. 
We express $\epsilon$ as a Fourier series in the Boozer angles,
\begin{align}
 \epsilon = \sum_{\deltam,\deltan} \epsilon_{\deltam,\deltan} e^{i(\deltam \theta - \deltan N_P\zeta)} = \sum_{\deltashort \overline{m},\deltan} \epsilon_{\deltashort \overline{m},\deltan} e^{i(\deltashort \overline{m} \chi - \deltan N_P\zeta)},
 \label{eq:Fourier_epsilon}
\end{align}
where $\deltam=\deltan=0$ terms are excluded according to the definition \eqref{eq:A_nae} and we have assumed field-period symmetry.
Here, we use the notation $\deltam$ and $\deltan$ to distinguish the equilibrium mode numbers from the mode numbers of the perturbed potential defined by \eqref{eq:SAW_continuum_Boozer}. Furthermore, this notation emphasizes that the equilibrium geometry will provide coupling between potential mode numbers $(m,n)$ and $(m+\deltam,n+\deltan)$. Since $\theta$ only enters through the angle $\chi$ in \eqref{eq:gradpsi2_nae}, it will sometimes be convenient to represent $\epsilon$ as a Fourier series in $(\chi,\zeta)$ as indicated in the above expression. When expressing $\epsilon$ as a Fourier series with respect to $\chi$, the distinction $\deltambar$ will be made apparent. We see from the near-axis expression \eqref{eq:A_nae} that, to leading order, only amplitudes $\epsilon_{\deltam,\deltan}$ with $\deltam = \pm 2$ or $\deltam = 0$ will be non-zero. 

\subsection{Rotating ellipse interpretation}

To gain further insight into the impact of shaping on the spectral content of $\epsilon$, we can relate averages of $\Psi_2$ to properties of the near-axis surfaces, which form ellipses in the plane perpendicular to the axis. Denoting the semi-major axis by $a$ and the semi-minor axis as $b$, the quantity $p = a^2 + b^2 = \left(1 + \overline{\kappa}^4(1 + \sigma^2)\right)/\overline{\kappa}^2$ controls the elongation $\mathcal{E} = a/b$ through the expression 
\begin{align}
    \mathcal{E} = \frac{p + \sqrt{p^2 - 4}}{2}.
        \label{eq:elongation}
\end{align}
Defining a coordinate system oriented with the ellipse axes, $x = a \cos \vartheta$ and $y = b \sin \vartheta$ for an ellipse parametrization angle $\vartheta$, the quantity $\Psi_2$ takes a particularly simple form:
\begin{align}
    \Psi_2 = B_0^2\frac{p - \sqrt{p^2 - 4} \cos(2\vartheta)}{2}.
    \label{eq:gradr2_ellipse}
\end{align}
We note that the flux surfaces become compressed around $\vartheta = \pi/2$, $3\pi/2$, corresponding to the semi-minor axis of the ellipse. Overall, the in-surface variation of the flux-surface compression increases with increasing elongation. 

To more precisely evaluate the harmonic content of $\Psi_2$, we must express the ellipse parametrization angle $\vartheta$ in terms of Boozer angles. It is the typical case of optimized quasisymmetric stellarators for their elliptical flux surfaces to make one half rotation with respect to the normal vector in one field period. While a solid theoretical justification for this feature is lacking (and would merit further exploration), there exists strong evidence for the persistence of this feature, as can be checked by analyzing standard QS designs like those in this paper, or more thoroughly, looking through the large database of near-axis configurations of \cite{2022Landreman_mapping}.\footnote{In fact, of all 448,743 QH fields in said database, $98.6\%$ exhibit half-rotation of their elliptical cross-section. Of the 74,387 QA configurations, 99.97\% do. This indicates a strong tendency to have a half-rotation, which can be interpreted in terms of a minimal amount of shaping that generates a sufficient amount of rotational transform \citep{mercier1964equilibrium}. However, there is no a priori reason why exceptions could not arise.} 
The direction of ellipse rotation (positive being counter-clockwise) will match the sign of $(\iota_0-N)$, being oppositely oriented for QA and QH configurations. (This can, for example, be seen from the Mercier formula, (67) in \citep{2020Jorgec}.)
For QH configurations, the normal vector makes one net rotation in the poloidal plane per field period, which counteracts the direction of ellipse rotation. On the other hand, for QA configurations, the normal vector does not make any net rotations. Therefore, for both QA and QH configurations, the elliptical surfaces make one half rotation in the poloidal plane per field period. Given this half-rotation of the ellipse with respect to the normal vector per field period, isocurves of constant $\vartheta$ will be helical when expressed in Boozer angles. We will assume that the sign of the toroidal angle is chosen so that the ellipse rotates in the positive direction in the poloidal plane, such that when expressed in terms of Boozer coordinates, $\theta - N_P \zeta$. 

Given \eqref{eq:gradr2_ellipse}, we therefore expect a strong helical $\deltam = 2$, $\deltan=1$ component of $\epsilon$, driven by rotating ellipticity. 
Additional toroidal coupling will be introduced due to the variation of the ellipticity parameter $p$ and rotation of the ellipse parametrization angle with respect to the Boozer angles, $\omega$ (which depends on $p$ and $\overline{\kappa}$). This could not only contribute to $(\deltam,\deltan) = (2,1)$, but also lead to the appearance of further couplings. These features are discussed further in Appendix \ref{app:NAE_ellipse}. 


\subsection{Boozer coordinate interpretation}

The spectral content of $\epsilon$ can now be explicitly evaluated through averages of \eqref{eq:gradpsi2_nae}, and can be expressed in terms of $\zeta$ averages, $\langle \dots \rangle_{\zeta} = (2\pi)^{-1}\int d\zeta \dots $, of the geometric parameters $p$, $\overline{\kappa}$, and $\gamma$:
\begin{align}
{\renewcommand{\arraystretch}{1.5}
  \left\{  
  \begin{array}{l}
  \epsilon_{\deltambar=0,n} = \frac{\left \langle p \cos(n N_P \zeta) \right \rangle_{\zeta}}{\left \langle  p \right \rangle_{\zeta}} \\
    \epsilon_{\deltambar=2,0} = \frac{\langle  \overline{\kappa}^2\rangle_{\zeta}}{\langle p \rangle_{\zeta}} - \frac{1}{2} \\
    \epsilon_{\deltambar=2,n} = \frac{\langle (2 \overline{\kappa}^2 - p) \cos(nN_P\zeta) - 2\tan(2\gamma)(2-p\overline{\kappa}^2) \sin(nN_P\zeta)\rangle_{\zeta}}{2\langle p \rangle_{\zeta}}.
    \end{array}
    \right.
}
\end{align}
Here, we have used the assumption of stellarator symmetry to deduce that $\Psi_2$ is even in $(m \chi - n N_P \zeta)$. As will be discussed in Section \ref{sec:perturbative_solution}, each of these spectral components of $\epsilon$ will drive a corresponding gap in the shear Alfv\'{e}n continuum, which we will label by the mode numbers $(\delta m,\delta n)$ , 
We will assume the standard language to describe these gaps: the TAE corresponds with $\delta m = 1$, $\delta n = 0$, the EAE gap corresponds with $\delta m = 2$, $\delta n = 0$, the HAE gaps corresponds with $\delta m \ne 0$ and $\delta n \ne 0$, and the MAE gaps corresponds with $\delta m=0$, $\delta n \ne 0$. 

We can now interpret the geometric origins of these spectral components and their corresponding gaps:
\begin{itemize}
    \item According to the discussion around \eqref{eq:gradr2_ellipse}, the $\epsilon_{\deltam=2,\deltan=1}$ spectral component is typically substantial in stellarators, driven by the rotating ellipse geometry, assuming that the sign of the toroidal angle is chosen to coincide with the direction of ellipse rotation. This component increases with increasing elongation through the parameter $p$. Thus rotating ellipticity drives the HAE (2,1) gap observed in many stellarators \citep{2001Kolesnichenko,2011Kolesnichenko}. 
    \item The mirror component, $\epsilon_{\deltambar=0,n}$, is produced by the toroidal variation of the elongation through the parameter $p$. Therefore, the toroidal variation of the elongation gives rise to the MAE gap. 
    \item $\epsilon_{\deltambar=2,0}$, which drives the elliptical component ($\deltam = 2$, $\deltan = 0$) of $\epsilon$ in QA configurations and the helical ($\deltam = 2$, $\deltan =2$) component in QH configurations, can be enhanced in the limit of either small or large $\langle 2\overline{\kappa}^2 \rangle_{\zeta}/\langle p \rangle_{\zeta}$ in comparison to 1. 
    In the limit of small and large $\overline{\kappa}$, the dominant balance ordering $\sigma \sim \overline{\kappa}^{-2}$ from \eqref{eq:sigma_equation} \citep{2023Rodriguezb} can be used to to obtain $p \rightarrow 2/\overline{\kappa}^2$ and $\overline{\kappa}^2$, respectively. Thus both limits of small or large $\langle 2\overline{\kappa}^2 \rangle_{\zeta}/\langle p \rangle_{\zeta}$ correspond to enhanced elongation. Therefore, elongation drives the EAE gap in QA configurations and the HAE (2,2) gap in QH configurations. 
    \item $\epsilon_{\deltambar=2,n}$, which drives the helical component ($\deltam = 2$, $\deltan = n$) in QA configurations and the helical ($\deltam = 2$, $\deltan = n + 2$) and elliptical components (for $n = -2$) in QH configurations, is produced by the toroidal variation of the curvature, elongation, and rotation angle. Thus the toroidal variation of the curvature and elongation drive other HAEs in QA and QH configurations and the EAE in QH configurations. 
\end{itemize}

\begin{figure}
    \centering 
    \includegraphics[width=0.9\textwidth]{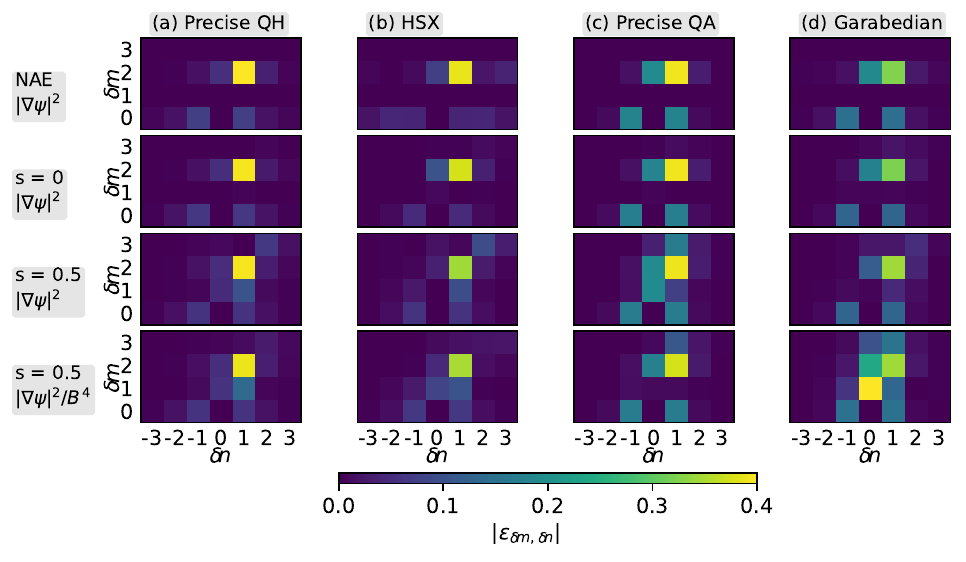}
    \caption{The Fourier transform of the normalized coupling parameters, $|\nabla \psi|^2/\langle |\nabla \psi|^2 \rangle$ and $|\nabla \psi|^2/B^4/\langle |\nabla \psi|^2/B^4\rangle$, are shown for two QH configurations, (a) and (b), and two QA configurations, (c) and (d). The near-axis expression $\Psi_2$, defined through \eqref{eq:psi_expansion}, (first row) shows good quantitative comparison with the on-axis equilibrium values (second row) and good qualitative comparison with the mid-flux value from the equilibrium (bottom row).} 
    \label{fig:FT_gradpsi2}
\end{figure}

In Figure \ref{fig:FT_gradpsi2}, the spectral content of the coupling parameters appearing in the continuum equation \eqref{eq:SAW_continuum_Boozer}, $|\nabla \psi|^2$ and $|\nabla \psi|^2/B^4$, are shown for two QH and two QA configurations \citep{2022Landreman,1995Anderson,2001Zarnstorff}.
Each equilibrium is fit to the near-axis representation \citep{2019Landreman}, and the corresponding expression for $|\nabla \psi|^2$ \eqref{eq:gradpsi2_nae} is Fourier transformed, as shown in the first row. In the second and third rows, the Fourier transform is performed on the $s = \psi/\psi_0 = 0$ and 0.5 surfaces of the equilibrium, where $\psi_0$ is the toroidal flux on the boundary. The spectral content of $|\nabla \psi|^2/B^4$ on the $s = 0.5$ surface is shown on the bottom row. 
For ease of interpretation, the sign of the toroidal angle is chosen such that the direction of ellipse rotation coincides for all equilibria.
Note that the spectral content of $|\nabla \psi|^2/B^4$ is equivalent to that of $|\nabla \psi|^2$ in the lowest-order near-axis expansion, so near-axis results are only shown for $|\nabla \psi|^2$.
We see that the near-axis expression provides quantitative agreement with the equilibrium on-axis, and qualitative agreement away from the axis.  
For all equilibria, the Fourier transform reveals that the dominant spectral component is the $\deltam = 2$, $\deltan = 1$ amplitude as predicted by the rotating ellipse geometry. 
Only for the QA equilibria do the elliptical and mirror components become comparable in magnitude, as these configuration have considerable toroidal variation of the elongation and curvature, see Table \ref{tab:A_comparison}.
The mirror and elliptical components are slightly larger in the Precise QA equilibrium than in the Garabedian equilibrium due to its enhanced elongation magnitude and variation. Away from the axis, the toroidal ($\deltam = 1$) and triangularity-induced ($\deltam = 3$) components are driven, which are not captured by the lowest-order near-axis theory.

To summarize, we note that the spectral content of $|\nabla \psi|^2$ is generally distinct from that of $B$. As discussed in previous work \citep{2001Kolesnichenko,2011Kolesnichenko}, due to the rotating ellipticity inherent to stellarator configurations near the magnetic axis, a strong helical $\deltam = 2$, $\deltan = \pm1$ component is driven, with a sign corresponding to the direction of ellipse rotation. Due to coupling through the toroidal variation of the curvature and elongation, additional mirror, helical, and elliptical components arise. This effect appears to be stronger in the QA configurations studied. While further study is required to determine if this trend holds over a wider database of configurations, a plausible explanation can be obtained from the fact that the normal vector of QH configurations must make $N$ rotations; thus the axis rotation naturally generates significant rotational transform. However, for QA configurations, more substantial elongation and shaping are required to generate substantial rotational transform. 




\begin{table}
    \centering
    \begin{tabular}{|c|c|c|c|c|c|c|c|c|c|}
       Configuration & $N$  & 
       Mean $\mathcal{E}$ & Var $\mathcal{E}$ & Mean $\overline{\kappa}$ & Var $\overline{\kappa}$ & $|\epsilon_{2,1}|$ & $|\epsilon_{2,0}|$ & $|\epsilon_{0,1}|$ \\ \hline 
        Precise QH & 4 & 2.9 & 0.33 & 1.2 & 1.1 & 0.80 & 0.11 & 0.15 \\
        HSX & 4 & 2.8 & 0.26 & 1.1 & 0.94 & 0.77 &  0.15 & 0.08  \\ 
        Precise QA  & 0 & 3.2 & 0.87 & 1.42 & 1.0  & 0.78 & 0.38 & 0.36 \\ 
        Garabedian & 0 & 2.4 & 0.81 & 2.9 & 0.8 & 0.65 & 0.38 & 0.29 \\ \hline
    \end{tabular}
    \caption{Configuration characteristics are shown for the equilibria in Figure \ref{fig:FT_gradpsi2} at $s = 0$, including the symmetry helicity $N$, mean and variance ($(\max(\mathcal{E})-\min(\mathcal{E}))/\text{mean}(\mathcal{E}$) of elongation, mean and variation of $\overline{\kappa}$, and the magnitudes of the dominant Fourier harmonics of $\epsilon_{\deltam,\deltan}$.}
    \label{tab:A_comparison}
\end{table}

\section{Perturbative solution to the continuum equation}
\label{sec:perturbative_solution}

Given the common physics basis for the formation of spectral gaps in the shear Alfv\'{e}n continuum in a torus and the band gaps in the electron energy spectrum in a periodic lattice \citep{2008Heidbrink}, it is natural to apply quantum mechanical techniques to analytically study the continuum structure in a near-axis quasisymmetric field. Here we use degenerate perturbation theory, treating the coupling parameter $\epsilon$ as a small perturbation, to identify the frequency splitting. Similar techniques have been applied to study the shear Alfv\'{e}n continuum of a large-aspect ratio tokamak \citep{1982Kieras,1986Riyopoulos}. This analysis will provide expressions for the spectral gaps in near-axis QS configurations, similar to previous studies of the continuum of model Helias configurations \citep{2001Kolesnichenko}. The formalism will also enable extensions beyond previous work, such as quantification of secondary gaps due to the interaction of multiple harmonics of $\epsilon$ and higher-order crossings, which have been observed in previous numerical studies \citep{2001Kolesnichenko,2007Yakovenko,2003Spong}.

\subsection{Near-axis perturbation theory}
\label{sec:perturbation_theory}

Given the continuum equation under the near-axis quasisymmetry model \eqref{eq:SAW_continuum_nae}, we now seek expressions for the continuum gap locations and width. Under the assumption that the coupling is small, $|\epsilon| \ll 1$, perturbation theory can be used to evaluate the eigenvalue shift associated with the coupling. To simplify the notation, we define the eigenvalue $\lambda_j = \overline{\omega}_j^2$. To lowest order in the coupling (denoted by superscript $(0)$), the continuum equation reads,
\begin{align}
    \left(\partder{}{\zeta} + \iota_0 \partder{}{\theta} \right) \left[
    \left(\partder{}{\zeta} + \iota_0 \partder{}{\theta} \right) \Phi^{(0)}_j\right] + \lambda_j^{(0)} \Phi_j^{(0)}  = 0.
\end{align}
The eigenfunctions are $\Phi^{(0)}_j = e^{i (m_j \theta - n_j \zeta + \omega t)}$ with eigenvalues 
\begin{align}
 \lambda_j^{(0)} = (\iota_0 m_j - n_j)^2. 
\label{eq:unperturbed_freq}
\end{align}

We continue to the linear order in the coupling parameter $\epsilon$, 
\begin{multline}
    \left(\partder{}{\zeta} + \iota_0 \partder{}{\theta} \right)^2 \Phi^{(1)} + \left[\left(\partder{}{\zeta} + \iota_0 \partder{}{\theta} \right) \epsilon \right]\left[\left(\partder{}{\zeta} + \iota_0 \partder{}{\theta} \right) \Phi^{(0)}  \right] \\
 + \lambda^{(0)} \Phi^{(1)} + \lambda^{(1)} \Phi^{(0)}  = 0.
  \label{eq:perturbed_continuum}
\end{multline}
We first consider non-degenerate perturbation theory, which assumes that all unperturbed frequencies are unique, before approaching the degenerate case. Non-degeneracy precludes the possibility of continuum frequency crossings, e.g. points where 
\begin{align}
    \iota_0 m_j - n_j = \pm \left(\iota_0 m_k - n_k \right)
    \label{eq:crossing_condition}
\end{align}
for some $(m_j,n_j) \ne (m_k, n_k)$. 
In this case, we assume \eqref{eq:perturbed_continuum} with $\Phi^{(0)} = \Phi_j^{(0)}$ and $\lambda^{(0)} = \lambda^{(0)}_j$. 
By integrating against $\left(\Phi^{(0)}_j\right)^*$ (star indicating the complex conjugate), noting that the operator $(\partial/\partial \zeta + \iota_0 \partial/\partial \theta)^2$ is self-adjoint, the constraint 
$\lambda_j^{(1)} = 0$
is obtained. Therefore no shift to the continuum frequency is obtained in the absence of degeneracy, i.e. continuum crossings. 

We next consider the case of two degenerate frequencies, $\lambda^{(0)} = \lambda^{(0)}_j = \lambda^{(0)}_k$, or equivalently \eqref{eq:crossing_condition}.
We define the mode number separations as $\Delta m = m_j - m_k$ and $\Delta n = n_j - n_k$. The two states are said to be co-propagating if they satisfy \eqref{eq:crossing_condition} with a positive sign, indicating the same sign of the parallel wave number $k_{\|}$ for both eigenfunctions. Otherwise, they are said to be counter-propagating. Here we adopt the notation $\overline{\omega}^{(0)}=  \sqrt{\lambda^{(0)}}$, with $\overline{\omega}_j^{(0)} =  \iota_0 m_j - n_j$. 
As will be shown below, only the counter-propagation case enables a frequency shift. In the counter-propagation case, the condition \eqref{eq:crossing_condition} implies 
\begin{align}
    \left \rvert \overline{\omega}^{(0)} \right \rvert =  \left \rvert \frac{\iota_0 \Delta m - \Delta n}{2} \right \rvert,
    \label{eq:gap_frequency}
\end{align}
while in the co-propagation case, it implies $ \iota_0 = \Delta n/\Delta m$. Therefore, the co-propagation case is only possible when multiple toroidal harmonics are considered, $\Delta n \ne 0$, as in the case of 3D configurations such as stellarators. 

\begin{figure}
\centering
\includegraphics[width=0.5\textwidth]{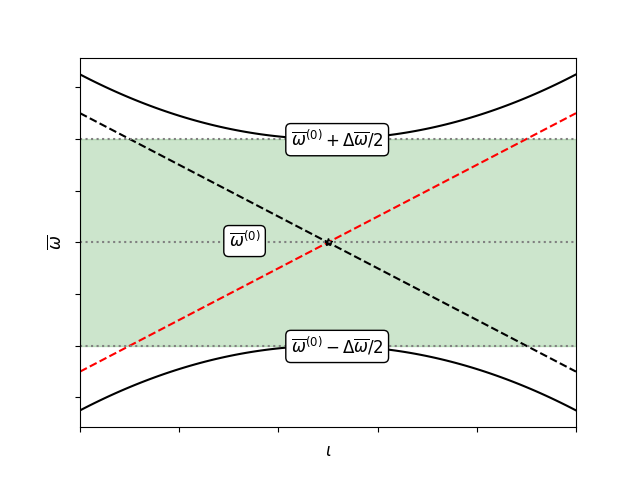}
\caption{A schematic diagram of a spectral gap formed due to a counter-propagating pair (dashed lines) that cross at $\omegazero$. Here red indicates $\omegazero>0$ and black indicates $\omegazero< 0$. In the presence of the coupling parameter $\epsilon$, a gap forms of width $\Delta \overline{\omega}$, given by \eqref{eq:gap_width}.}
\label{fig:simple_schematic}
\end{figure}

In the case of two-way degeneracy of either sign, the unperturbed eigenfunction is a
linear combination of two unperturbed states,
\begin{align}
    \Phi^{(0)} = \alpha_j\Phi^{(0)}_j + \alpha_k \Phi^{(0)}_k,
\end{align}
for unknown amplitudes $\alpha_j$ and $\alpha_k$ with $\lambda^{(0)}_j = \lambda^{(0)}_k = \lambda^{(0)}$.
The perturbed continuum equation \eqref{eq:perturbed_continuum}
is then integrated against $\left(\Phi^{(0)}_j\right)^*$ and $\left(\Phi^{(0)}_k\right)^*$ to yield a set of coupled equations for the amplitudes $\alpha_j$ and $\alpha_k$
\begin{align}
    \left[ \begin{array}{cc}
\lambda^{(1)}  & \epsilon_{\Delta m,\Delta n} \overline{\omega}^{(0)}_j \left(\iota_0 \Delta m - \Delta n\right) \\
 \epsilon_{\Delta m,\Delta n}^* \overline{\omega}^{(0)}_j \left(\iota_0 \Delta m - \Delta n\right)          & \lambda^{(1)} 
    \end{array} \right]\left[ \begin{array}{c}\alpha_j \\ \alpha_k \end{array}\right]
    = \left[\begin{array}{c} 0 \\ 0 \end{array}\right].
\end{align}
Setting the determinant of the above matrix to zero provides the frequency shift,
\begin{align}
\left(\lambda^{(1)}\right)^2 =|\epsilon_{\Delta m, \Delta n/N_P}|^2 \left(\iota_0 \Delta m - \Delta n \right)^2 \lambda^{(0)},
\end{align}
noting that $\left(\overline{\omega}_j^{(0)}\right)^2 = \lambda^{(0)}$. Here, we have assumed $N_P$-symmetry, with the convention that the toroidal mode number of $\epsilon$ is multiplied by $N_P$ in the definition \eqref{eq:Fourier_epsilon}.
In the case of co-propagation, there is evidently no frequency shift, and the two modes will continue to cross each other unless a higher-order degeneracy is present as described in Section \ref{sec:higher_degeneracy}. For the counter-propagating case, the frequency shift is evaluated from 
the displacement of the positive and negative solutions, $\Delta \overline{\omega} = \sqrt{\lambda^{(0)} + \lambda^{(1)}_+} - \sqrt{\lambda^{(0)} + \lambda^{(1)}_-}$, approximated as
\begin{align}
\Delta \overline{\omega} = 2\overline{\omega}^{(0)} |\epsilon_{\Delta m,\Delta n/N_P}|.
\label{eq:gap_width}
\end{align}
Therefore, while in the absence of coupling, the frequencies would cross at the point indicated by \eqref{eq:gap_frequency}, the crossing is avoided in the presence of coupling. The phenomena of avoided crossings is often referred to as a spectral gap associated with mode numbers $\Delta m$ and $\Delta n/N_P$. See Figure \ref{fig:simple_schematic} for a schematic diagram. 
We will refer to such locations as the ($\Delta m,\Delta n/N_P$) gap. As described in Section \ref{sec:away_from_axis}, the avoided crossings persist away from the axis, with the central gap frequency $\overline{\omega}^{(0)}$ being a continuous function of $\iota$. 

In an axisymmetric system, all toroidal mode numbers are decoupled, since the geometric factor $\epsilon$ is independent of $\zeta$. Thus one can analyze the continuum independently for each toroidal mode number $n$, and $\Delta n = 0$ for all crossings. In this way, all crossings are avoided if coupling is present, since co-propagation is not possible. However, in a 3D system, not all crossings are avoided due to co-propagation described above and gap crossings, as described in Section \ref{sec:higher_order_crossing}. This additional complexity generally obscures spectral gaps in 3D systems, although one may still gain basic intuition from this perturbative analysis. 

The following Sections will further extend this theory. Section \ref{sec:higher_degeneracy} will consider the case when the degeneracy is not lifted at the linear order, but at quadratic order. Section \ref{sec:higher_order_crossing} will consider the case of higher-order crossings, which are permitted in 3D systems. Finally, Section \ref{sec:away_from_axis} will extend the perturbation analysis away from the magnetic axis. 


\subsection{Higher-order degeneracy}
\label{sec:higher_degeneracy}

We now consider the case when the degeneracy is not lifted at first-order in $\epsilon$, implying $\epsilon_{\Delta m,\Delta n/N_P} = \lambda^{(1)} = 0$. Even in the absence of a frequency shift at first order, there may still be a shift to the eigenfunction, $\Phi^{(1)}$, which enters at $\mathcal{O}(\epsilon^2)$. Since the lowest-order eigenfunctions are a complete basis, the correction to the eigenfunction can be expressed as,
\begin{align}
   \Phi^{(1)} = \sum_{i\ne{j,k}} \mu_i \Phi_i^{(0)},
   \label{eq:phi_1}
\end{align}
where we are free to assume that $\Phi^{(1)}$ has no projection onto the degenerate subspace, $\Phi_j^{(0)}$ and $\Phi_k^{(0)}$. The coefficients $\mu_i$ are determined by integrating \eqref{eq:perturbed_continuum} against $\left(\Phi^{(0)}_i\right)^*$,
\begin{align}
    \mu_i = \frac{\Delta_{ij}\epsilon_{ij}\alpha_j\overline{\omega}_j + \Delta_{ik}\epsilon_{ik}\alpha_k\overline{\omega}_k}{\lambda^{(0)}-\lambda_i^{(0)}},
    \label{eq:mu_i}
\end{align}
where $\Delta_{ij} = \iota_0 \Delta m_{ij} - \Delta n_{ij}$ and $\epsilon_{ij} = \epsilon_{\Delta m_{ij}, \Delta n_{ij}/N_P}$. 
We now continue to second order in $\epsilon$:
\begin{multline}
    \left(\partder{}{\zeta} + \iota_0 \partder{}{\theta} \right)^2 \Phi^{(2)} +  
    \left[\left(\partder{}{\zeta} + \iota_0 \partder{}{\theta} \right) \epsilon \right]\left[\left(\partder{}{\zeta} + \iota_0 \partder{}{\theta} \right) \Phi^{(1)}  \right] \\
    - \epsilon \left[\left(\partder{}{\zeta} + \iota_0 \partder{}{\theta} \right) \epsilon \right]\left[\left(\partder{}{\zeta} + \iota_0 \partder{}{\theta} \right) \Phi^{(0)}  \right]  \\
  + \lambda^{(0)}  \Phi^{(2)} + \lambda^{(2)} \Phi^{(0)}  = 0.
  \label{eq:2nd_order_degenerate_continuum}
\end{multline}
Integrating against $\left(\Phi_j^{(0)}\right)^*$ and $\left(\Phi_k^{(0)}\right)^*$, using \eqref{eq:phi_1} and \eqref{eq:mu_i} again results in a set of coupled equations for $\alpha_j$ and $\alpha_k$. 
The zero-determinant condition, assuming counter-propagation, then reads:
\begin{align}
        \left(\frac{\lambda^{(2)}}{\lambda^{(0)}} + F\right)^2 - |E|^2 = 0,
\end{align}
with,
\begin{align}
E &= \sum_{\deltam,\deltan}\epsilon_{\deltam,\deltan} \epsilon_{\Deltam-\deltam,\Deltan/N_P-\deltan}, \\
F &= \sum_{\deltam,\deltan}  |\epsilon_{\deltam,\deltan}|^2 \frac{\Delta({\deltam,\deltan})}{\Delta({\deltam,\deltan}) + 2 \overline{\omega}^{(0)}_j},
\end{align}
where $\Delta({\deltam,\deltan}) = \iota_0 \deltam-\deltan$. The gap width due to two-harmonic coupling then reads:
\begin{align}
\Delta \overline{\omega} = \overline{\omega}^{(0)}|E| .
\label{eq:gap_width}
\end{align}
In addition to the frequency splitting, the gap central frequency is modified to $\overline{\omega}_j^{(0)}\sqrt{1-F}$.
See Appendix \ref{app:second_order_degeneracy} for details. 

Because the frequency shift is second-order in the coupling parameters, such gaps will typically be narrower. This behavior has been observed in numerical continuum calculations, sometimes referred to as secondary or higher-order gaps \citep{2001Kolesnichenko}. As an example, according to near-axis theory, $\epsilon$ can only provide harmonics with $\deltam = 2$ or 0. However, a gap with $\Delta m = 4$ can nonetheless arise due to higher-order degeneracy, as will be demonstrated numerically in Section \ref{sec:numerical_continuum}. More generally, $n$-order degeneracies may also arise, whose width scales with $\epsilon^n$.


\subsection{Higher-order crossings}
\label{sec:higher_order_crossing}

 While spectral gaps remain well-separated in 2D configurations, they may generally cross when $\Delta n \ne 0$. The phenomena of gap crossing results in higher-order crossing of unperturbed frequencies, which can reduce the gap width and modify the gap structure. We first remark on the interpretation of such higher-order crossings before embarking on the analysis. 
 
 If multiple pairs of modes are counter-propagating, this implies the intersection of spectral gaps with different mode numbers, $(\Delta m_1,\Delta n_1)$ and $(\Delta m_2,\Delta n_2)$. Since the frequency center of the gap is given by \eqref{eq:gap_frequency}, the crossing of these spectral gaps implies 
\begin{align}
    \iota_0 = \frac{\Delta n_1 - \Delta n_2}{\Delta m_1 - \Delta m_2}. 
\end{align}
Thus the gap crossing necessarily occurs on a rational surface. If one gap crossing exists, satisfying the above condition, then there are an infinite number of gap crossings at the same location since the numerator and denominator can be scaled by the same factor. For example, gaps labeled by mode numbers ($\Delta m_1$, $\Delta n_1$) and ($\Delta m_3$, $\Delta n_3$) with $\Delta n_3 = k\Delta n_2 - (k-1)\Delta n_1$ and $\Delta m_3 = k \Delta m_2 - (k-1) \Delta m_1$ will cross at the same location for any integer $k$. Thus gap crossing implies many-way crossings at the intersection point. As we will see in Section \ref{sec:numerics_crossing}, with increasing numerical resolution, the number of crossings increases. As illustrative cases, we discuss 3-way and 4-way crossings below. 

As discussed in \citep{2007Yakovenko}, at the gap crossing point, the continuum equation \eqref{eq:2nd_order_degenerate_continuum} then becomes decoupled across field lines since only parallel derivatives appear. When the two gaps cross, the parallel wavenumber associated with the two corresponding coupling parameters, $\epsilon_{\Delta m_1,\Delta n_1/N_P}$ and $\epsilon_{\Delta m_2, \Delta n_2/N_P}$, then coincide, since $\iota_0 \Delta m_1 - \Delta n_1 = \pm (\iota_0 \Delta m_2 - \Delta n_2)$. In this way, $\epsilon_{\Delta m_1,\Delta n_1/N_P}$ and $\epsilon_{\Delta m_2, \Delta n_2/N_P}$ can enable coupling between both pairs of counter-propagating modes. The variation across field lines of the coupling can act to reduce the gap width, with the resulting width being approximately the difference in the widths of the crossing gaps. This behavior will be demonstrated in Section \ref{sec:numerics_crossing}.

Now we consider the situation of multiple of pairs of modes co-propagating with each other, implying that 
\begin{align}
    \iota_0 = \frac{\Delta n_1}{\Delta m_1} = \frac{\Delta n_2}{\Delta m_2}. 
\end{align}
Again, this condition can only be satisfied on a rational surface, and co-propagation can only occur for helical gaps ($\Delta m_1$, $\Delta n_1$, $\Delta m_2$, $\Delta n_2$ are all non-zero), Evidently, the two pairs of modes must have a common factor. Thus this behavior occurs due to intersection with a higher harmonic of a helical gap (e.g., ($\Delta m$, $\Delta n$) = (2,1) and (4,2)). As with the counter-propagating case, an infinite number of gap crossings is possible by rescaling the numerator and denominator by the same integer. 

We now analyze the case of higher-order crossings at linear order. Since the behavior will differ for odd- and even-numbered crossings, we will discuss the examples of three-way and four-way crossings.

As an illustrative case, we first consider three degenerate frequencies, $\Phi^{(0)} = \alpha_i \Phi_i^{(0)} + \alpha_j \Phi_j^{(0)} + \alpha_k \Phi_k^{(0)}$. We define the mode number separations as $\Delta m_{i,j} = m_i - m_j$ and $\Delta n_{i,j} = n_i-n_j$. Following a similar argument to Section \ref{sec:perturbation_theory}, 
the linear system determining the frequency shift is
\begin{align}
    \left[ \begin{array}{c c c} 
    \lambda^{(1)} & -\epsilon_{ij} \Delta_{ij} \overline{\omega}^{(0)}_j & -\epsilon_{ik} \Delta_{ik}\overline{\omega}^{(0)}_k \\
    \epsilon_{ij}^* \Delta_{ij} \overline{\omega}^{(0)}_i & \lambda^{(1)} & - \epsilon_{jk}\Delta_{jk}\overline{\omega}^{(0)}_k \\
    \epsilon_{ik}^*\Delta_{ik}\overline{\omega}^{(0)}_i & \epsilon_{jk}^*\Delta_{jk}\overline{\omega}^{(0)}_j & \lambda^{(1)}
    \end{array}\right] \left [ \begin{array}{c} \alpha_i \\
    \alpha_j \\ \alpha_k 
    \end{array} \right] = \left [ \begin{array}{c} 0 \\
    0 \\ 0
    \end{array} \right], 
\end{align}
where $\Delta_{ij} = \iota_0 \Delta m_{ij} - \Delta n_{ij}$ and $\epsilon_{ij} = \epsilon_{\Delta m_{ij}, \Delta n_{ij}/N_P}$. 
Without loss of generality, we can assume that either all three unperturbed eigenfunctions are co-propagating, $\overline{\omega}^{(0)}_i = \overline{\omega}^{(0)}_j = \overline{\omega}^{(0)}_k$, or two are co-propagating and the third is counter-propagating $\overline{\omega}^{(0)}_i = -\overline{\omega}^{(0)}_j = \overline{\omega}^{(0)}_k$. In the purely co-propagation case, the zero-determinant condition then reads:
\begin{align}
    \left(\lambda^{(1)}\right)^3 = 0, 
\end{align}
while in the case with counter-propagation, 
\begin{align}
\lambda^{(1)} \left[\left(\lambda^{(1)}\right)^2 - 4 \left(\lambda^{(0)} \right)^2 \left(|\epsilon_{jk}|^2 + |\epsilon_{ij}|^2\right) \right] = 0. 
\end{align}
In either case, there remains at least one solution with no frequency shift, $\lambda^{(1)} = 0$. In the case with counter-propagation, there remains two solutions with a shifted frequency depending on the two harmonics of $\epsilon$ which couple a pair of counter-propagating modes: 
\begin{align}
\Delta \overline{\omega} = 2\overline{\omega}^{(0)} \sqrt{|\epsilon_{jk}|^2 + |\epsilon_{ij}|^2}.
\label{eq:freq_shift_3x3}
\end{align}
However, at least one continuum eigenfunction will not be shifted in frequency and will appear to ``cross the gap.'' A similar structure persists for higher-order crossings of an odd number. Since co-propagation cannot be avoided for odd crossings, not all crossings will be avoided. 

\begin{figure}
    \centering 
    \begin{subfigure}{0.49\textwidth}
    \includegraphics[width=0.99\textwidth]{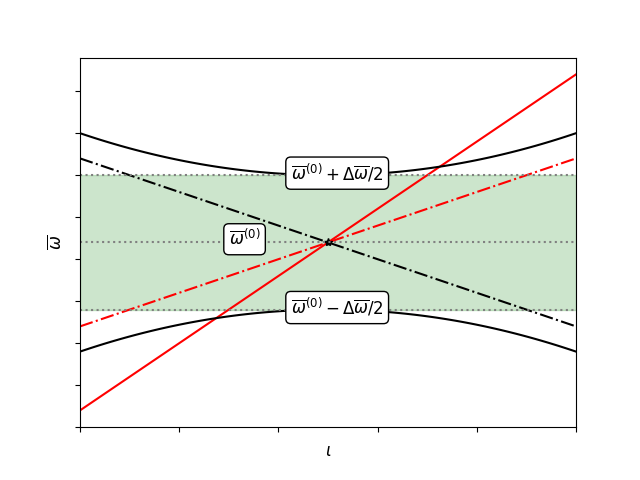}
    \caption{}
    \end{subfigure}
    \begin{subfigure}{0.49\textwidth}
    \includegraphics[width=0.99\textwidth]{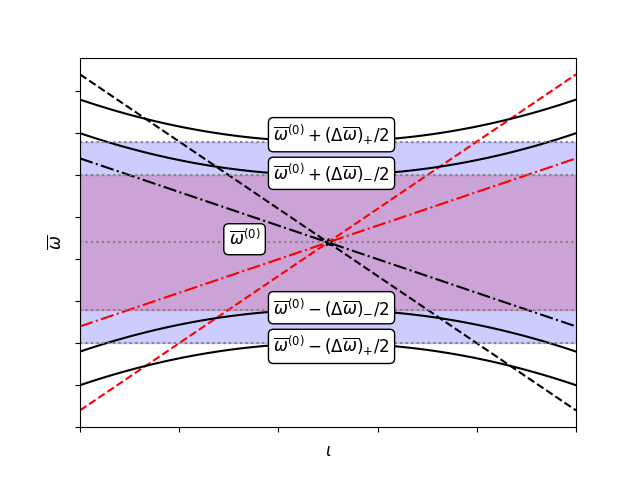}
    \caption{}
    \end{subfigure}
    \caption{Schematic diagrams of higher-order crossings. Here red indicates $\omegazero>0$ and black indicates $\omegazero< 0$. In (a), there is a three-way crossing at $\omegazero$. The counter-propagating pair (dashed lines) is shifted by the perturbation, forming the gap indicated by the green shaded region. The solid red line is unshifted by the perturbation and appears to cross the gap. In (b), there is a four-way crossing at $\omegazero$. The two counter-propagating pairs are both shifted by the perturbation, forming gaps of different widths, indicated by the shaded regions. The effective gap, where continuum damping is minimized, is the region of overlap between the two gaps. }
    \label{fig:multiple_crossing_diagram}
\end{figure}

To consider higher-order crossings of an even number, we now consider the case of a four-way degeneracy, $\Phi^{(0)} = \alpha_i \Phi_i^{(0)} + \alpha_j \Phi_j^{(0)} + \alpha_k \Phi_k^{(0)} + \alpha_l \Phi_l^{(0)}$. 
Without loss of generality, we can consider three possibilities: all modes are co-propagating ($\overline{\omega}^{(0)}_i = \overline{\omega}^{(0)}_j = \overline{\omega}^{(0)}_k = \overline{\omega}^{(0)}_l$), three are co-propagating and one is counter-propagating with the others ($\overline{\omega}^{(0)}_i = -\overline{\omega}^{(0)}_j = -\overline{\omega}^{(0)}_k = -\overline{\omega}^{(0)}_l$), and two pairs are counter-propagating ($\omegazero_i = -\omegazero_j = \omegazero_k = - \omegazero_l$). In the first case, the zero-determinant condition reads $\left(\lambda^{(1)} \right)^4 = 0$, and again, no frequency shift results. In the case of three co-propagating modes, the frequency shifts satisfy
\begin{align}
    \left(\lambda^{(1)} \right)^2 \left[\left(\lambda^{(1)} \right)^2 - 4 \left(\lambda^{(0)}\right)^2 \left(|\epsilon_{ij}|^2 + |\epsilon_{ik}|^2 + |\epsilon_{il}|^2 \right)\right] = 0. 
\end{align}
Here, two of the frequencies are shifted according to harmonics of $\epsilon$ which couple two counter-propagating modes, analogous to \eqref{eq:freq_shift_3x3}:
\begin{align}
    \Delta \overline{\omega} = 2\omegazero\sqrt{|\epsilon_{ij}|^2 + |\epsilon_{ik}|^2 + |\epsilon_{il}|^2},
\end{align}
while the other two frequencies are not shifted by the perturbation. Finally, we consider the case of two pairs of counter-propagating modes:
\begin{multline}
\left(\lambda^{(1)}\right)^2 = 4  \left(\lambda^{(0)}\right)^2 \Bigg(\left[|\epsilon_{ij}|^2 + |\epsilon_{il}|^2 + |\epsilon_{jk}|^2 + |\epsilon_{kl}|^2\right] \\ \pm \sqrt{\left(|\epsilon_{ij}|^2 + |\epsilon_{il}|^2 + |\epsilon_{jk}|^2 + |\epsilon_{kl}|^2\right)^2 - 4|\epsilon_{il} \epsilon_{jk}^* - \epsilon_{ij} \epsilon_{kl}|^2}\Bigg), 
\end{multline}
and all solutions enable frequency shift by harmonics of $\epsilon$ that couple counter-propagating modes. Note that there are now two solutions for the gap width, and each counter-propagating pair is shifted by a distinct width: 
\begin{multline}
    \left(\Delta \overline{\omega}\right)_{\pm} = 2 \omegazero \Big(\left[|\epsilon_{ij}|^2 + |\epsilon_{il}|^2 + |\epsilon_{jk}|^2 + |\epsilon_{kl}|^2\right] \\ \pm \sqrt{\left(|\epsilon_{ij}|^2 + |\epsilon_{il}|^2 + |\epsilon_{jk}|^2 + |\epsilon_{kl}|^2\right)^2 - 4|\epsilon_{il} \epsilon_{jk}^* - \epsilon_{ij} \epsilon_{kl}|^2}\Big)^{1/2}. 
\end{multline}
The effective gap width will be the smaller of these two solutions, since this is the region in which continuum damping can be avoided. In the limit that only $\epsilon_{ij}$ and $\epsilon_{kl}$ are non-zero, enabling coupling between these two counter-propagating pairs, the gap width solutions reduce to $\Delta \overline{\omega} = 2 \overline{\omega}^{(0)}|\epsilon_{ij}|$, $2 \overline{\omega}^{(0)}|\epsilon_{kl}|$, as is consistent with \eqref{eq:gap_width}. See Figure \ref{fig:multiple_crossing_diagram} for a schematic diagram. 

To summarize, higher-order crossings are allowed in 3D configurations due to coupling between modes with $\Delta m \ne 0$ and $\Delta n \ne 0$. If all of the degenerate eigenfunctions are co-propagating, then no frequency shift is present. If some modes are counter-propagating, then there will exist a frequency shift related to the harmonics of $\epsilon$ which couple the counter-propagating modes. For an odd-numbered crossing, at least one degenerate mode frequency will not be shifted, while for an even-numbered crossing, it is possible for all frequencies to be shifted by the perturbation if every eigenfunction is in a counter-propagating pair. This behavior will be shown in numerical continuum solutions in Section \ref{sec:numerics_crossing}.

\subsection{Behavior away from magnetic axis}
\label{sec:away_from_axis}

In this Section, we relax the near-axis assumption to evaluate behavior away from the axis. From \eqref{eq:SAW_continuum_Boozer}, two coupling parameters now arise in the continuum equation :
\begin{align}
    \left(\partder{}{\zeta} + \iota \partder{}{\theta} \right) \left[(1 + \epsilon_1)
    \left(\partder{}{\zeta} + \iota \partder{}{\theta} \right) \Phi\right] + \overline{\omega}^2  (1 + \epsilon_2) \Phi  = 0,
\end{align}
with
\begin{align}
   \left \{ \begin{array}{l} \epsilon_1 = \frac{|\nabla \psi|^2}{\langle |\nabla \psi|^2\rangle } -1 \\
    \epsilon_2 = \frac{|\nabla \psi|^2/B^4}{\langle |\nabla \psi|^2/B^4 \rangle} - 1 \end{array} \right. 
\label{eq:def_eps_1_2}
\end{align}
and $\overline{\omega} = \omega/\omega_A$, defined by $\omega_A = \langle B\rangle^2/(G+\iota I)\sqrt{\mu_0 \rho})$. A similar perturbative analysis can be performed as described in Section \ref{sec:perturbation_theory}, where now the two coupling parameters are assumed to be small, $|\epsilon_1| \ll 1$ and $|\epsilon_2| \ll 1$. The unperturbed frequencies are then given by 
\begin{align}
    \left \rvert \overline{\omega}^{(0)} \right \rvert =  \left \rvert \frac{\iota \Delta m - \Delta n}{2} \right \rvert. 
    \label{eq:gap_frequency_away_axis}
\end{align}
Again, at linear order, a frequency shift only arises in the degenerate counter-propagating case, given by
\begin{align}
    \Delta \overline{\omega} = \omegazero \sqrt{|\epsilon_1^{\Delta m,\Delta n/N_P}|^2 + |\epsilon_2^{\Delta m,\Delta n/N_P}|^2},
\end{align}
where $\epsilon_{1,2}^{\Delta m,\Delta n/N_P}$ are defined analogously to \eqref{eq:Fourier_epsilon}. In this way, the region of avoided crossings persists away from the axis, with the central gap frequency depending on the rotational transform through \eqref{eq:gap_frequency_away_axis}.

The coupling parameters can now be estimated by proceeding to next order in the near-axis expansion:
\begin{align}
\renewcommand*{\arraystretch}{1.5}
    \left \{ \begin{array}{l} |\nabla \psi|^2/r^2 = \Psi_2 + \Psi_3 r + \mathcal{O}(r^2) \\
    |\nabla \psi|^2/(r^2 B^4) = \Psi_2 (1 - 4 r \overline{\eta} \cos(\chi)) + r \Psi_3 +\mathcal{O}(r^2) \end{array} \right. .
\end{align}

According to the discussion in Appendix \ref{app:third_order_nablapsi2}, the geometric parameter $\Psi_3$ is driven by higher-order shaping components, such as triangularity and the Shafronov shift. With the $\cos(\theta)$ and $\cos(3\theta)$ dependence of $\Psi_3$, the TAE ($\Delta m =1$) and noncircular-triangularity induced Alfvén eigenmodes (NAE) ($\Delta m = 3$)  \citep{1998Kramer} gaps are driven. These same gaps are also driven through the beating of the $\cos(2\chi)$ dependence of $\Psi_2$ and the $\cos(\chi)$ in the above expression. Through numerical examples in Section \ref{sec:optimization}, we will see the formation of the TAE and NAE in continuum solutions away from the magnetic axis. 

\subsection{Summary}

To summarize, through perturbative analysis under the assumption of smallness of the coupling parameter in the continuum equation, frequency shifts of counter-propagating continuum modes are computed. This enables the identification of the location \eqref{eq:gap_frequency} and width \eqref{eq:gap_width} of spectral gaps and their relation to the near-axis quasisymmetric geometry described in Section \ref{sec:nae_qs}. Even if counter-propagation enables the formation of a continuum gap, co-propagation remains a possibility for helical gaps, for which $\Delta m \ne 0$ and $\Delta n \ne 0$. This may result in continuum modes which appear to ``cross the gap.'' The theory has been extended to evaluate cases for which the degeneracy is not lifted at first order in $\epsilon$, resulting in the phenomena of gaps formed due to coupling of different harmonics of $\epsilon$. We also assess higher-order frequency crossings that may exist in 3D configurations. In the case of an odd-numbered crossing, there remains at least one frequency that is unshifted and may appear to cross through the gap. We find that several coupling parameters may contribute to the frequency width in the case of such higher-order crossings. Finally, we extend the theory away from the magnetic axis to account for higher-order contributions of the flux-surface shaping and field strength variation. 


\section{Numerical continuum calculations}
\label{sec:numerical_continuum}

We now validate the predictions based on perturbation theory in Section \ref{sec:perturbation_theory} through numerical solutions of \eqref{eq:SAW_continuum_nae} for near-axis configurations of interest. The continuum equation is solved using a Fourier Galerkin method, similar to the STELLGAP code \citep{2003Spong}. The expression for the coupling parameter \eqref{eq:A_nae} is evaluated using pyQSC \citep{pyQSC} for near-axis configurations fitted to known equilibria. Since it is not possible to discern gap locations by computing the continuum at a single radial point (since an avoided crossing is likely not to occur at that radial grid point), we evaluate the continuum for a uniform grid corresponding to the range of $\iota$ of interest. The range of $\iota$ is chosen to be large enough that a sufficient number of crossings and avoided crossings can be visualized. Given the low magnetic shear of typical optimized stellarators, the range of $\iota$ will be relatively small for the following calculations. 
The geometric parameter $\epsilon$ is computed from the same leading order coupling, $\Psi_2$, for the range of $\iota$. Note that for all of the following calculations, the normalized frequency, $\overline{\omega} = \omega/\omega_A$, is shown, a quantity that is independent of the choice of density profile. 

\subsection{Fourier mode number choice}

For numerical efficiency, we choose our set of Fourier basis functions in order to provide sufficient resolution of the low-frequency behavior of interest \citep{2003Spong,1999Nuhrenberg}. Since \eqref{eq:unperturbed_freq} provides an approximate relation between the mode numbers and frequency of a continuum eigenfunction, the set of poloidal and toroidal mode numbers $m$ and $n$ included in the spectrum can be chosen strategically. 
In order to resolve the frequency range $\overline{\omega}_0 \in [-|\overline{\omega}_0|_{\max},+|\overline{\omega}_0|_{\max}]$, for a given range of $m$, the set of $n$ between $-|\overline{\omega}_0|_{\max} + \iota_{\min}m$ and $|\overline{\omega}_0|_{\max} + \iota_{\max}m$ are included, where $\iota_{\min}$ and $\iota_{\max}$ are the minimum and maximum values of the rotational transform to be studied. 
All toroidal mode numbers $n$ are chosen to belong to the same mode family: a set of toroidal mode numbers that differ by integer multiples of $N_P$. (For example, the $n = 0$ mode family contains the toroidal mode numbers $0, \pm N_P, \pm 2 N_P$, etc.) The resolution parameter $|\overline{\omega}_0|_{\max}$ is adjusted to ensure the frequency range of interest is resolved. With this choice of Fourier modes, one can more efficiently resolve the eigenmode structure than by using the same range of $n$ for all $m$. Furthermore, by reducing the range of frequencies resolved, the condition number of the discretized problem is reduced. 

On the other hand, the choice of $m_{\max}$ could impact the behavior of the continuum within the frequency range of interest. While the qualitative features of the continuum should remain with increasing $m_{\max}$, true convergence cannot be expected. When adding $m$ modes, additional eigenfunctions are expected to be present in the low-frequency region, enabling additional crossings and avoided crossings. While avoided crossings of the same type are expected to persist within the same gap region, there may arise continuum-crossing modes due to an odd-numbered crossing, as described in Section \ref{sec:higher_order_crossing}. Thus when adjusting this parameter, we do not expect the precise structure of the continuum to remain unchanged. However, the qualitative features, such as the characteristic width between avoided crossings, should be retained with increasing resolution. Further discussion of numerical aspects of solving the continuum equation will be provided in a follow-up paper. 

\subsection{Gap width validation}

\begin{figure}
    \centering 
    \includegraphics[width=0.5\textwidth]{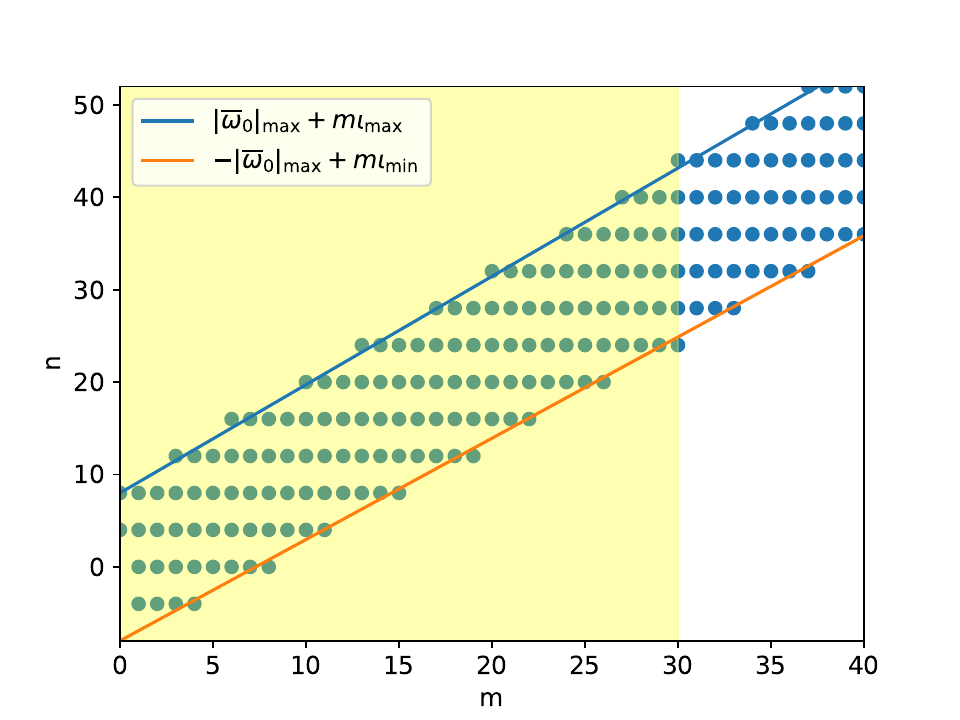}
    \caption{The continuum is computed for a near-axis Wistell-A configuration \citep{2020Bader} using the Fourier spectral basis with mode numbers indicated in the figure, using the mode choice scheme with $m_{\max} = 40$. The yellow shaded region corresponds with the set of modes include in the calculations labeled $m_{\max} = 30$ in Figure \ref{fig:aten}.}
    \label{fig:aten_mode_choice}
\end{figure}

We now validate the gap width presented in Section \ref{sec:perturbation_theory} using a near-axis QH configuration fit to the Wistell-A equilibrium \citep{2020Bader}. The continuum equation \eqref{eq:SAW_continuum_nae} is solved using the near-axis geometry, but with the full range of $\iota$ in the equilibria in order to more clearly visualize the continuum structure. The modes, visualized in Figure \ref{fig:aten_mode_choice} as blue dots, are selected using $|\overline{\omega}_0|_{\max} = 2 N_P$, $m_{\max} = 30$ and 40, and mode family 0. When $m_{\max} = 30$, only modes within the yellow shaded box are included, while all visualized modes are included when $m_{\max} = 40$. 

\begin{figure}
    \centering 
    \begin{subfigure}{0.49\textwidth}
    \centering 
    \includegraphics[width=0.99\textwidth]{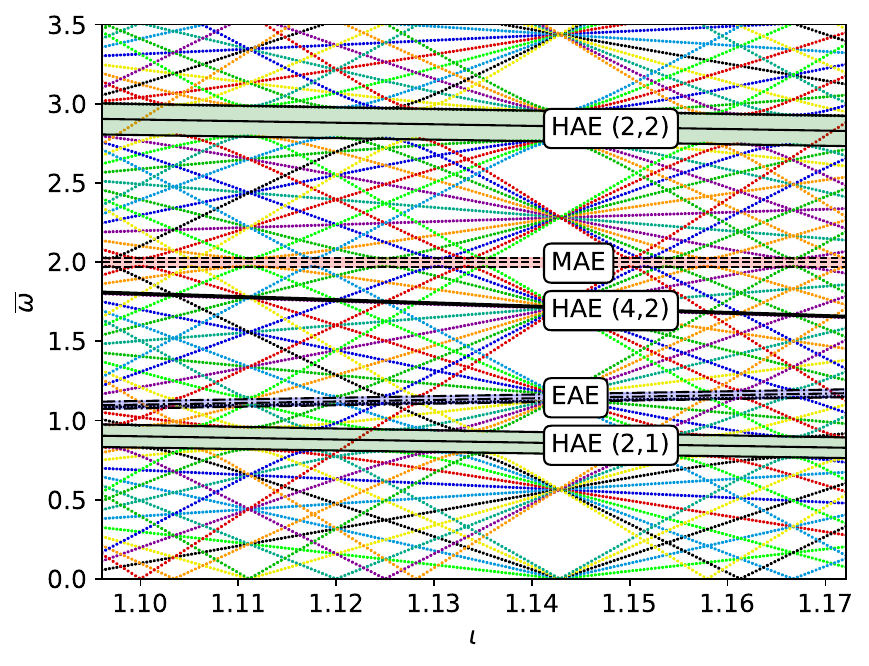}
    \caption{$\epsilon$ scaled by 0.25, $m_{\max} = 40$}
    \end{subfigure}
    \begin{subfigure}{0.49\textwidth}
    \centering 
    \includegraphics[width=0.99\textwidth]{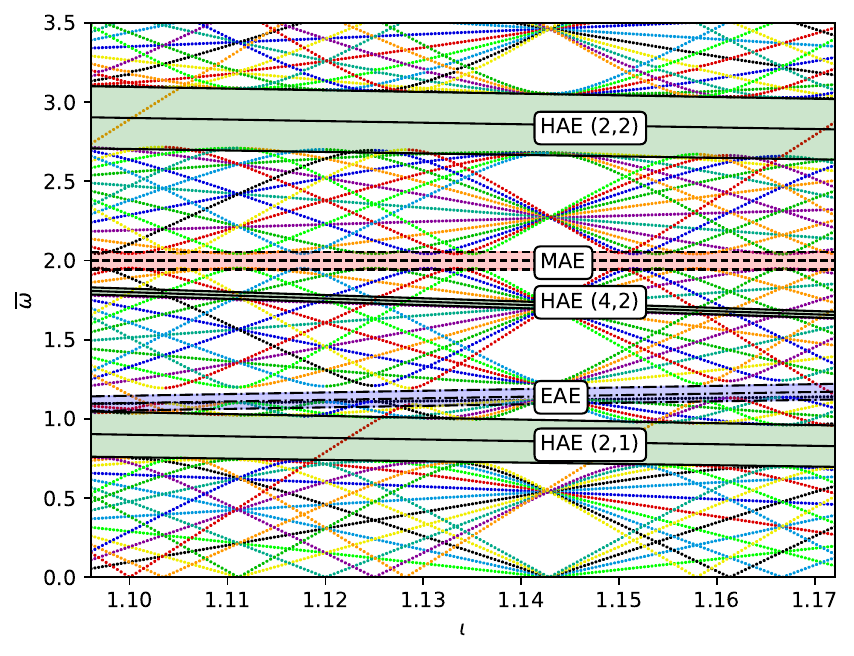}
    \caption{$\epsilon$ scaled by 0.5, $m_{\max} = 40$}
    \label{fig:aten_eps_0.5_mmax_40}
    \end{subfigure}
    \begin{subfigure}{0.49\textwidth}
    \centering 
    \includegraphics[width=0.99\textwidth]{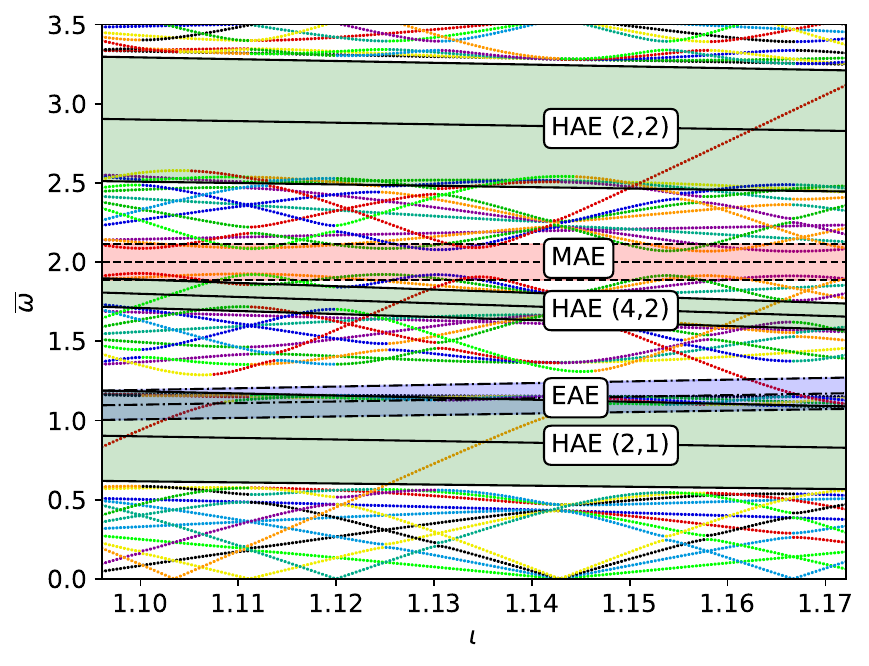}
    \caption{$\epsilon$ unscaled, $m_{\max} = 30$}
    \label{fig:aten_eps_1.0_mmax_30}
    \end{subfigure}
    \begin{subfigure}{0.49\textwidth}
    \centering 
    \includegraphics[width=0.99\textwidth]{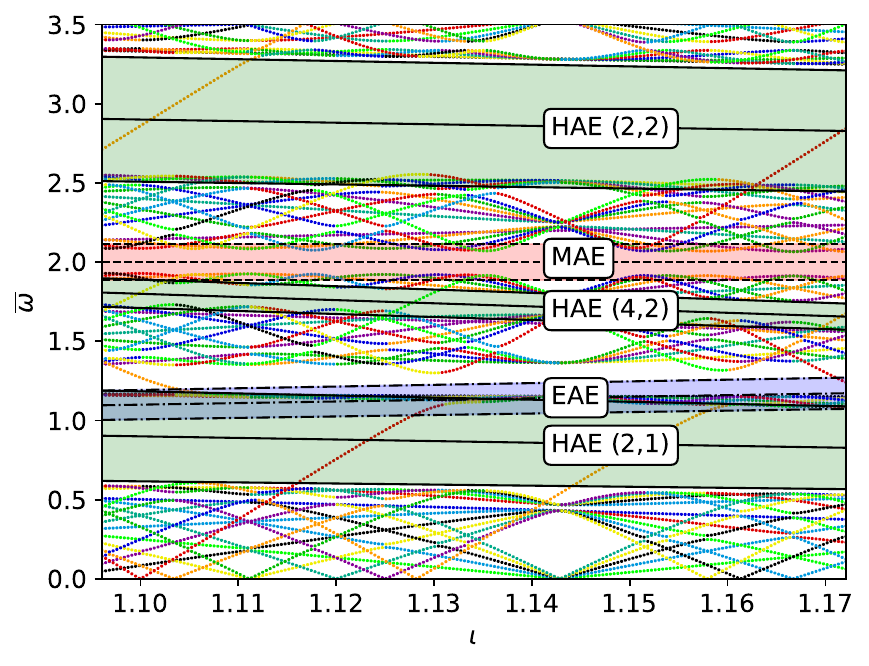} 
    \caption{$\epsilon$ unscaled, $m_{\max} = 40$}
    \label{fig:aten_eps_1.0_mmax_40}
    \end{subfigure}
    \caption{The continuum is computed for a near-axis Wistell-A configuration \citep{2020Bader} using the Fourier spectral basis with mode numbers indicated in Figure \ref{fig:aten_mode_choice}. In each of the continuum figures, the color scale indicates the dominant poloidal mode number of the eigenfunction while the colored shaded regions indicate the predicted spectral gaps.}
    \label{fig:aten}
\end{figure}

The computed continuum eigenmode frequencies are shown in Figure \ref{fig:aten}. Here the color scale indicates the dominant poloidal mode number of the corresponding eigenfunction while the shaded colored regions indicate the theoretically-predicted gaps. Since the theory was developed in the small $\epsilon$ limit, the continuum is shown with $\epsilon$ scaled by a constant factor of 0.25 and 0.5 in addition to the unscaled value. There is good agreement between the continuum gaps and predicted gaps, especially for smaller $\epsilon$. Note that the HAE (4,2) gap arises due to coupling between the $\deltam = 2, \deltan = 1$ harmonics of $\epsilon$ according to Section \ref{sec:higher_degeneracy}, since the near-axis spectral content of $\epsilon$ does not contain $\deltam = 4$ harmonics. As $\epsilon$ is increased, the EAE and HAE (2,1) gaps begin to cross with each other, and the EAE gap is displaced away from its predicted position. This is indicative of gap repulsion discussed in the literature \citep{2001Kolesnichenko}. While continuum gaps are apparent, we note the presence of eigenmodes which cross the gaps. For example, there is an eigenmode with dominant Fourier mode numbers $n = 32$, $m = 20$ mode that crosses the HAE (2,1) gap near $\iota = 1.135$. When the resolution is increased to $m_{\max} = 40$, this eigenmode couples with an eigenmode dominated by $n = 36$, $m = 31$ to avoid crossing. However, in its place an $n = 44$, $m = 40$ eigenmode crosses the gap. This behavior highlights that such continuum-crossing modes are an artifact of resolution choice and should not inhibit the identification of spectral gaps. 

\subsection{Validation of higher-order crossings}
\label{sec:numerics_crossing}

\begin{figure}
\centering 
\begin{subfigure}{0.49\textwidth}
\centering
\includegraphics[width=0.99\textwidth]{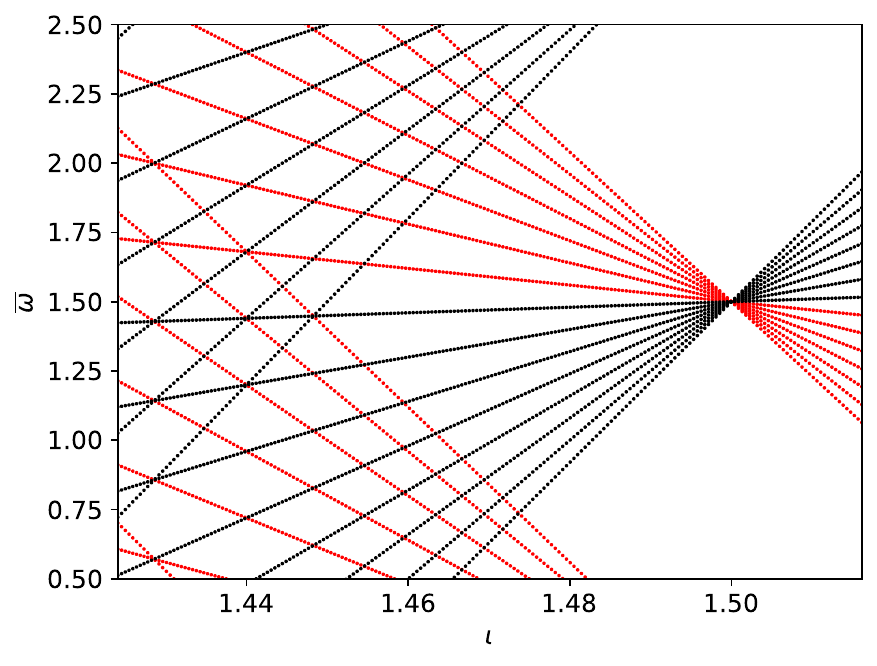}
\caption{$\epsilon = 0$, $m_{\max} = 30$}
\label{fig:eps_0_mmax_30}
\end{subfigure}
\begin{subfigure}{0.49\textwidth}
\centering
\includegraphics[width=0.99\textwidth]{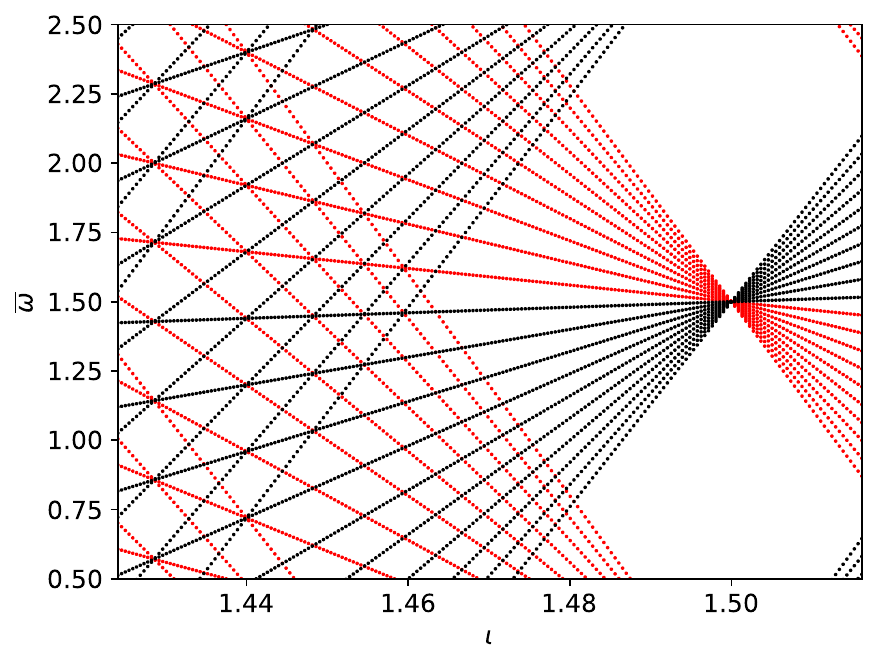}
\caption{$\epsilon = 0$, $m_{\max} = 40$}
\label{fig:eps_0_mmax_40}
\end{subfigure}
\begin{subfigure}{0.49\textwidth}
\centering
\includegraphics[width=0.99\textwidth]{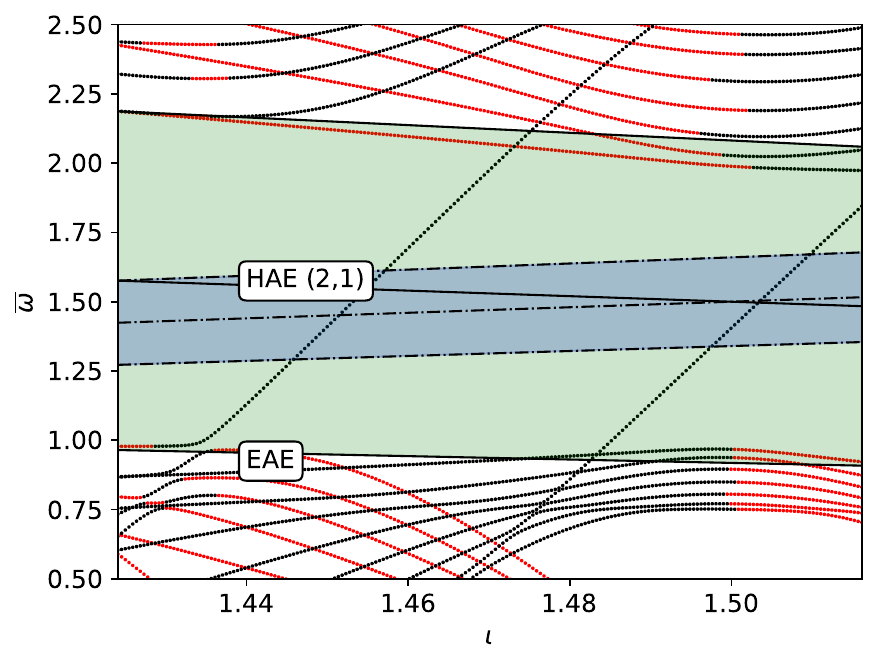}
\caption{unscaled $\epsilon$, $m_{\max} = 30$}
\label{fig:eps_1_mmax_30}
\end{subfigure}
\begin{subfigure}{0.49\textwidth}
\centering
\includegraphics[width=0.99\textwidth]{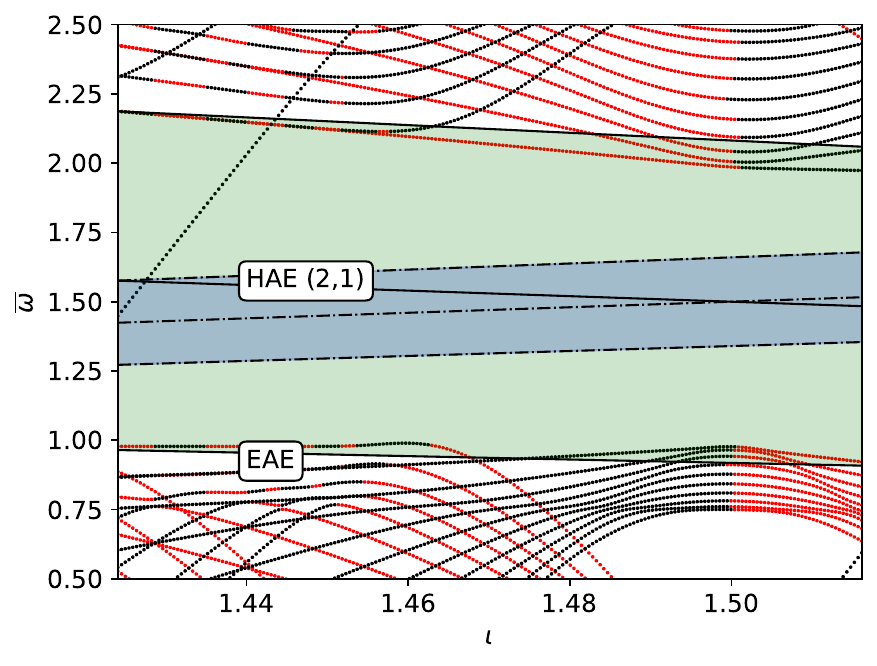}
\caption{unscaled $\epsilon$, $m_{\max} = 40$}
\label{fig:eps_1_mmax_40}
\end{subfigure}
\caption{The continuum is computed for a near-axis N\"{u}hrenberg-Zille configuration \citep{1988Nuhrenberg}. In each of the continuum figures, the color scale indicates the direction of eigenmode propagation: red indicates $\overline{\omega}>0$ and black indicates $\overline{\omega}<0$. The colored shaded regions indicates the predicted spectral gaps.}
\end{figure}

In order to evaluate higher-order crossings, we compute the continuum for a near-axis configuration with parameters obtained from the N\"{u}hrenberg and Zille QH configuration \citep{1988Nuhrenberg}. The mode choice parameters $m_{\max} = 30$ and 40, $|\overline{\omega}_0|_{\max} = 2 N_P$, and mode family 0 are used. 
The continuum solution without coupling ($\epsilon = 0$, physically corresponding to the cylindrical limit) is shown in Figures \ref{fig:eps_0_mmax_30} and \ref{fig:eps_0_mmax_40}. Here the color scale indicates the direction of propagation: red indicates $\overline{\omega}>0$ while black indicates $\overline{\omega}<0$. A high-order crossing is evident at $\iota = 1.5$ and $\overline{\omega} = 1.5$. This is the location for the intersection of the EAE and HAE (2,1) gaps. The continuum solutions for unscaled $\epsilon$ are shown in Figures \ref{fig:eps_1_mmax_30} and \ref{fig:eps_1_mmax_40}, with predicted gap locations indicated by the colored shaded regions. Due to the interaction between the two gaps, the EAE gap is repelled away from the HAE gap, and the width of the HAE (2,1) gap is reduced at the gap intersection point. Note that as the resolution parameter $m_{\max}$ is increased from 30 to 40, the crossing changes from odd numbered (15-way) to even-numbered (20-way). As expected based on the discussion in Section \ref{sec:higher_order_crossing}, in the odd-numbered crossing case, there remains one continuum-crossing eigenmode, while the remaining eigenmodes couple in counter-propagating pairs to avoid crossing the gap. In the even-numbered crossing case, all eigenmodes couple in counter-propagating pairs, and no continuum crossing mode remains at this location. In this case, the counter-propagating pairs form nested gaps of different widths as anticipated. However, another continuum-crossing eigenmode arises around $\iota = 1.43$ due to the behavior discussed in the previous section in relation to Figure \ref{fig:aten}. Again, the presence of a continuum-crossing mode can be considered an artifact of the choice of Fourier mode numbers and should not inhibit the identification of a gap. 

\subsection{Analysis of selected configurations}
\label{sec:selected_configs}

We now numerically compute the shear Alv\'{e}n continuum for the four near-axis configurations presented in Section \ref{sec:nae_qs}. The mode choice parameters $m_{\max} = 60$, $|\overline{\omega}_0|_{\max} = 2 N_P$, and mode family 0 are used. For the HSX and Garabedian equilibria, the continuum is computed for the full range of rotational transform. In the case of the Precise QA and Precise QH equilibria, the range of $\iota$ is extended beyond the values in the equilibria in order to more clearly visualize the gap structure, given their low shear. 

Although the four configurations have similar magnitudes of the $\epsilon_{2,1}$ component (see Table \ref{tab:A_comparison}), we note that the width of the corresponding HAE (2,1) gap is significantly larger in the QH configurations. This can be explained by the scaling of the gap width \eqref{eq:gap_width} with the central frequency \eqref{eq:gap_frequency}, which is increased for QH configurations in comparison to QA configurations given their larger rotational transform and number of field periods. In both configurations, higher harmonics of the HAE (2,1) are also excited (e.g., HAE (4,2) and HAE (6,3)) due to the higher-order degeneracy effect discussed in Section \ref{sec:higher_degeneracy}, given the large magnitude of $\epsilon_{2, 1}$. This effect is more prominent in the QH configurations, given the larger central frequency of these gaps. 

The elliptical $\epsilon_{2,0}$ and mirror $\epsilon_{0,1}$ spectral components are more substantial in the QA configurations than the QH configurations, given their toroidal variation of the elongation and curvature. However, the EAE and MAE gaps are visible in all configurations due to scaling of the gap width with the central frequency. The EAE central gap frequency is magnified in the larger $\iota$ QH configurations. Similarly, the central frequency of the MAE gap is magnified in the QH configurations given their larger values of $N_P$. The implications of these general trends for resonance with energetic particles will be discussed in Section \ref{sec:gap_resonance}. In a follow-up paper, we will extend this comparison to the SAW continuum computed from optimized stellarator equilibria rather than a near-axis model.

\begin{figure}
\centering
\begin{subfigure}{0.49\textwidth}
\includegraphics[width=0.99\textwidth]{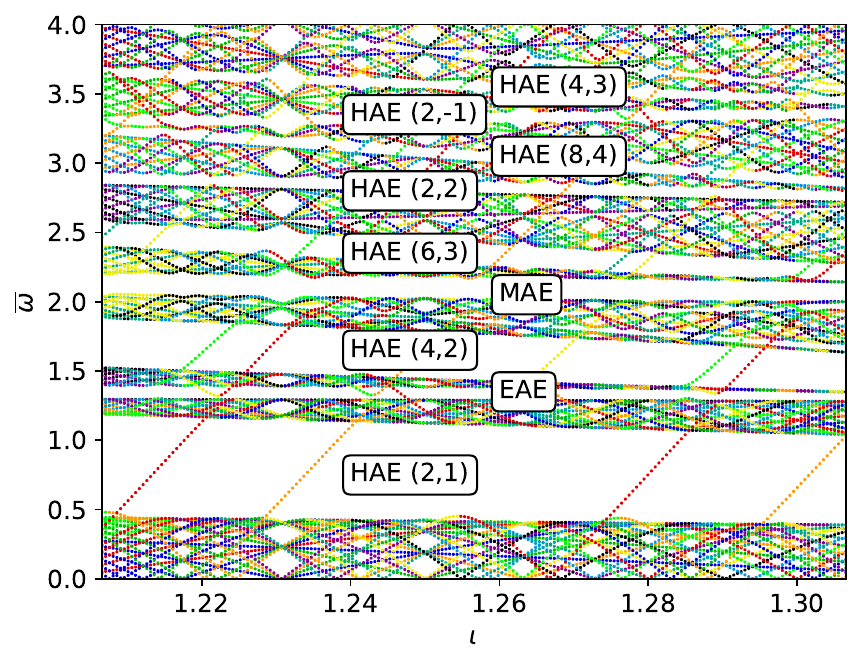}
\caption{Precise QH}
\end{subfigure}
\begin{subfigure}{0.49\textwidth}
\includegraphics[width=0.99\textwidth]{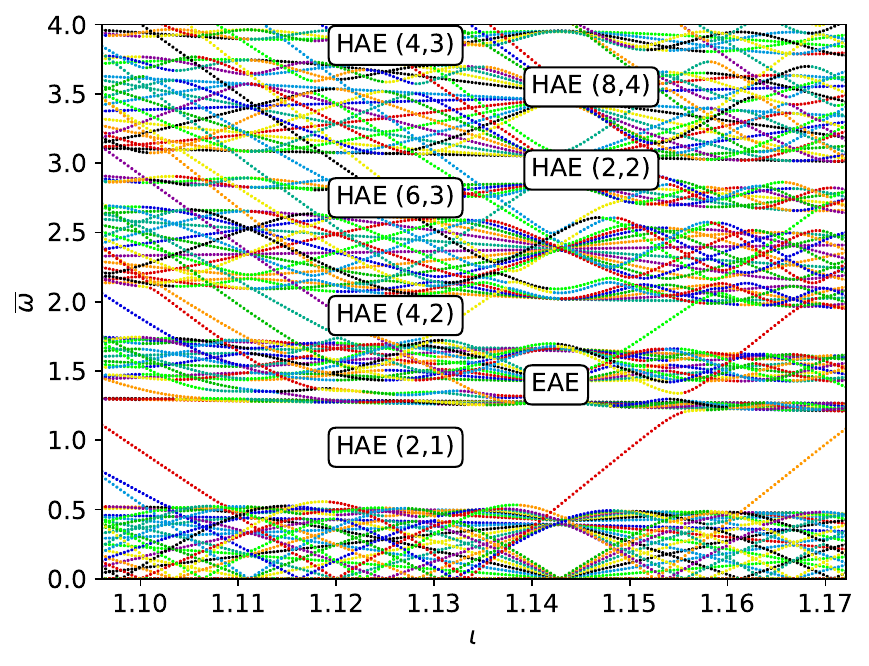}
\caption{HSX}
\end{subfigure}
\begin{subfigure}{0.49\textwidth}
\includegraphics[width=0.99\textwidth]{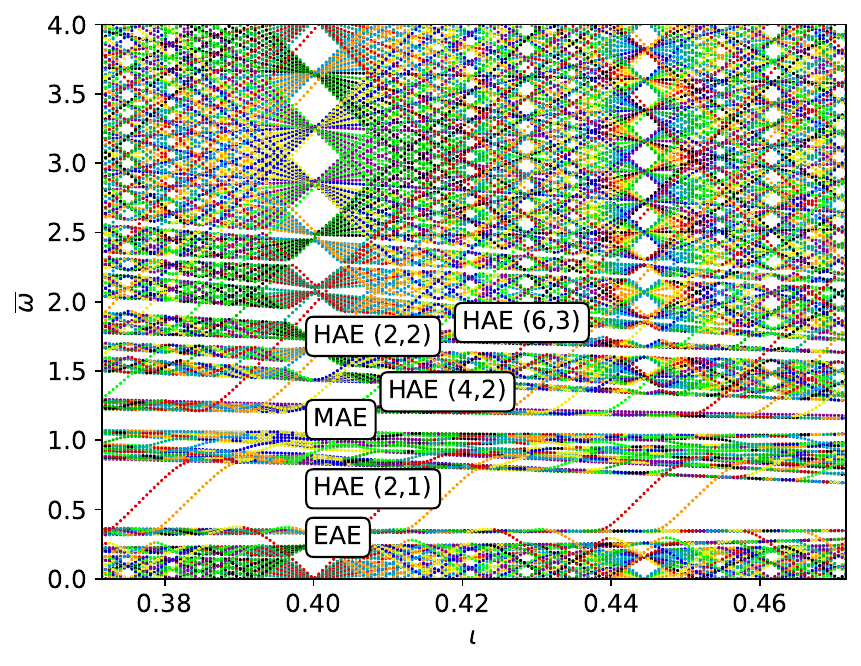}
\caption{Precise QA}
\end{subfigure}
\begin{subfigure}{0.49\textwidth}
\includegraphics[width=0.99\textwidth]{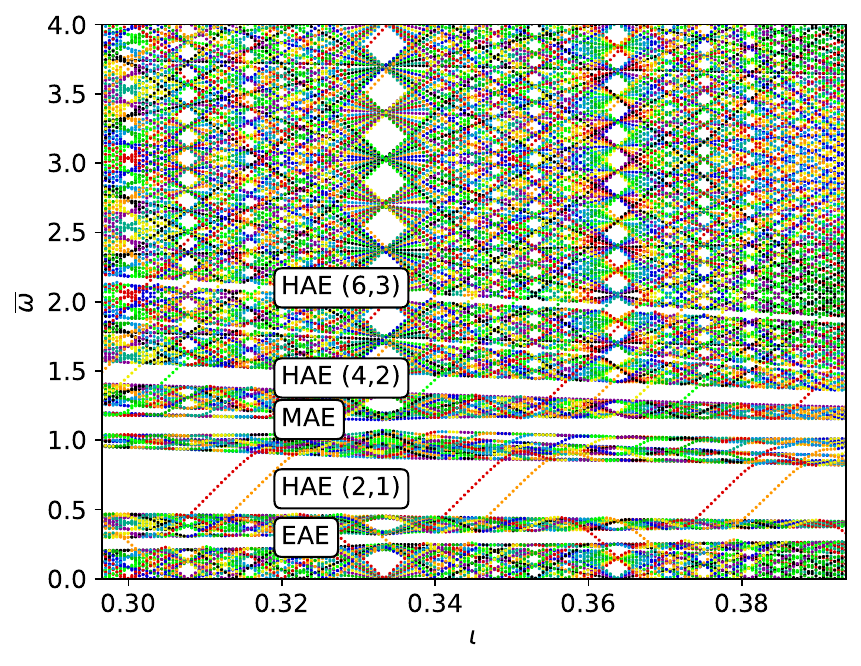}
\caption{Garabedian}
\end{subfigure}
\caption{The continuum is evaluated for the four near-axis configurations discussed in Section \ref{sec:nae_qs}. Here the color scale indicates the dominant poloidal mode number of the eigenfunction. The dominant spectral gaps are labeled based on visual inspection of the frequency interactions.}
\label{fig:continuum_select_figs}
\end{figure}




\section{Resonance condition for a gap Alfv\'{e}n eigenmode}
\label{sec:gap_resonance}

Given the formation of continuum gaps, there is potential for global Alfv\'{e}n eigenmodes described by \eqref{eq:SAW} to be driven unstable by resonant interaction with energetic particles. Such AE gap modes are typically dominated by the continuum mode numbers which couple to produce the gap \citep{1986Cheng,Betti1991}. We now evaluate the passing alpha particle resonance condition to assess the potential for instability of gap modes in quasisymmetric devices. To avoid strong alpha transport, one might desire to suppress continuum gaps which could resonate with alphas from birth to thermalization. Since this would be quite geometrically restrictive, and since prompt losses are the most harmful, we first focus on the resonance condition at the birth energy. 



The resonance condition for passing particles in the presence of a SAW with mode numbers $m$ and $n$ and frequency $\omega$ reads
\citep{2023Paul},
\begin{align}
    (m + l)\omega_{\theta} - (n + lN)\omega_{\zeta} + \omega  = 0,
\end{align}
where $l$ is a parameter that denotes sideband coupling through the drifts. Typically the resonance is strongest for $l = 0$ or $\pm 1$. Here $\omega_{\theta}$ and $\omega_{\zeta}$ are the averaged precession frequencies in the $\theta$ and $\zeta$ directions. For simplicity, we consider the case of co- or counter-passing particles, for which $\omega_{\theta}/\omega_{\zeta} \approx \pm \iota$. Assuming that the dominant modes numbers of the gap AE will correspond with the mode numbers of the degenerate continuum eigenfunctions, the counter-propagation condition \eqref{eq:crossing_condition} is used to obtain  $\iota m - n = \pm (\iota \Delta m - \Delta n)/2$. 
The AE frequency is also approximated by the central gap frequency, $\omega \approx \pm \omega_A (\iota \Delta m - \Delta n)/2$ to obtain the resonance condition for co-passing particles:
\begin{align}
   \left \rvert \frac{\omega_A}{\omega_{\zeta}} \right \rvert = \left \rvert 1 \pm \frac{2l(\iota - N)}{\iota \Delta m - \Delta n}\right \rvert,
   \label{eq:resonance_AE}
\end{align}
and counter-passing particles:
\begin{align}
    \left \rvert \frac{\omega_A}{\omega_{\zeta}} \right \rvert = \left \rvert 1 \pm \frac{4n + 2l(\iota + N)}{\iota \Delta m - \Delta n} \right \rvert.
\end{align}

Taking parameters of the ARIES-CS \citep{2008Najmabadi} QA reactor study ($B = 5.86$ T, $n_i = 4.8 \times 10^{20}$ m$^{-3}$) or the HSR418 Helias study \citep{2001Beidler} ($B = 5$ T, $n_i = 2.6 \times 10^{20}$ m$^{-3}$), the Alfv\'{e}n velocity is expected to be about $v_A \approx 3 \times 10^6$ m/s in a stellarator reactor, compared to the alpha birth velocity of $1.3 \times 10^7$ m/s. 
Since $\omega_A/\omega_{\zeta}$ will likely be smaller than 1, the resonance condition will be easiest to satisfy for $|l| = 1$ co-passing particles. (It is more challenging to satisfy the counter-passing condition, given the large mode numbers expected at FPP, $n \sim 30$ \citep{2014Gorelenkov}.) The condition $\omega_A/\omega_{\zeta} < 1$ with $|l| = 1$ implies,
\begin{align}
    |\iota - N| < |\iota \Delta m - \Delta n|.
    \label{eq:passing_res}
\end{align}
We now assess the potential for satisfying this condition in quasisymmetric devices. First, we consider the case of QA configurations, for which $N = 0$ and typically $\iota \lesssim 0.5$ \citep{2019Landreman}.
It is plausible to satisfy the resonance condition for HAE modes with $\Delta m \ge 1$ and $\Delta n \ge N_P$, MAE modes with $\Delta n \ge N_P$, and EAE modes with $\Delta m = 2$ and $\Delta n =0$. TAE modes with $\Delta m = 1$ and $\Delta n = 0$ are less likely to satisfy the resonance condition. Overall, the passing resonance is challenging to avoid in QA configurations.

Next, we consider the case of QH configurations, for which $N \approx 4-5$ and $\iota \approx 1-1.5$ \citep{2019Landreman}.
Satisfying the resonance condition would requires a gap mode with $\Delta m \ge 3$ and $\Delta n = 0$, an MAE mode with $\Delta n \ge N_P$, or some HAE modes with $\Delta m > 0$, $\Delta n > 2 N_P$. However, it is challenging to satisfy this resonance condition for EAE or the $(2,1)$ HAE mode. 

In summary, there are several avenues to avoiding passing resonances. First, high-density operation reduces $\omega_A$, making the passing resonance condition more challenging to achieve. Furthermore, it appears more challenging to satisfy the gap resonance condition in QH configurations, since resonance with modes residing in the wide EAE or HAE (2,1) gaps are weaker. Finally, as discussed in the next Section, the flux surface geometry can be manipulated to avoid strong resonances associated with high-frequency gaps. We remark that as the gaps are moved to lower frequency, there may be modification to the shear Alfv\'{e}n continuum due to the sound wave coupling, since $v_A/c_s \approx 0.3$, where $c_s$ is the sound speed, using the above mentioned reactor parameters with $T= 10$ keV. Furthermore, there may be resonances with alphas as they slow down, but such induced transport will be less deleterious.

\section{A pathway toward continuum optimization}
\label{sec:optimization}

Given the immense freedom in the stellarator design space, we now discuss a potential design criterion to reduce resonance with gap AE modes. Because of the close connection between AE gaps and flux-surface geometry, this is a promising pathway toward optimization of stellarators for improved AE stability. While it is likely impossible to eliminate all spectral gaps, since this would require $|\nabla \psi|^2$ to be a constant function on magnetic surfaces, the spectral gaps can be strategically manipulated to avoid strong resonances that would drive prompt losses near the birth energy. Since the passing resonance condition with $\omega_A/\omega_{\zeta} < 1$ requires sufficiently large values of the normalized frequency at the center of the gap, $\omega/\omega_A \approx (\iota \Delta m - \Delta n)/2$, one can attempt to close the gaps associated with high-frequency modes. A further motivation to promote low-frequency continuum gaps arises because the gap width is proportional to the gap frequency, as can be seen in \eqref{eq:gap_frequency}.

We perform fixed-boundary optimization of the vacuum QH Wistell-A equilibrium \citep{2020Bader} with the following objective function depending on the plasma boundary $S_P$:
\begin{align}
    f(S_P) = \left(A(S_P)-A^*\right)^2 + f_{QS}(S_P) + f_{\iota}(S_P) + f_{\text{cont}}(S_P). 
\end{align}
Here $A(S_P)$ is the aspect ratio and $A^* = 6.7$ is the target aspect ratio (same as the initial equilibrium). The function $f_{QS}$ is the quasisymmetry error \citep{2022Rodriguezb,2022Landreman}: 
\begin{align}
    f_{QS} = \sum_s \left \langle \left(\frac{1}{B^3} \left[(N-\iota) \bm{B} \times \nabla B \cdot \nabla \psi - (G + NI) \bm{B}\cdot \nabla B\right] \right)^2   \right \rangle, 
\end{align}
where $N = 4$ is the quasisymmetry helicity. The function $f_{\iota}$ prevents the rotational transform from getting too close to the $\iota = 1$ resonance \citep{2022Landremanb}:
\begin{align}
    f_{\iota} = \sum_s  |\min(|\iota|-1.03,0)|^2. 
\end{align}
The function $f_{\text{cont}}$ prevents the formation of high-frequency gaps that can satisfy the resonance condition \eqref{eq:passing_res}:
\begin{align}
    f_{\text{cont}} = \sum_s \sum_{|\iota \deltam - N_P\deltan| > |\iota - 4|}
|\epsilon_1^{\deltam,\deltan}|^2 \left(\iota \deltam - N_P \delta n \right)^2.
\end{align}
The gap width \eqref{eq:gap_width} associated with mode numbers ($\deltam$,$\deltan$) is proportional to $\overline{\omega}^{(0)}|\epsilon_1^{\deltam,\deltan}|^2$, where the central gap frequency is $\overline{\omega}^{(0)} = |\iota \deltam - N_P\deltan|/2$. Thus the objective function quantifies the squared gap width for each ($\deltam$,$\deltan$) that could satisfy the resonance condition. 

\begin{figure}
    \centering
    \begin{subfigure}{0.49\textwidth}
\includegraphics[width=0.99\textwidth]{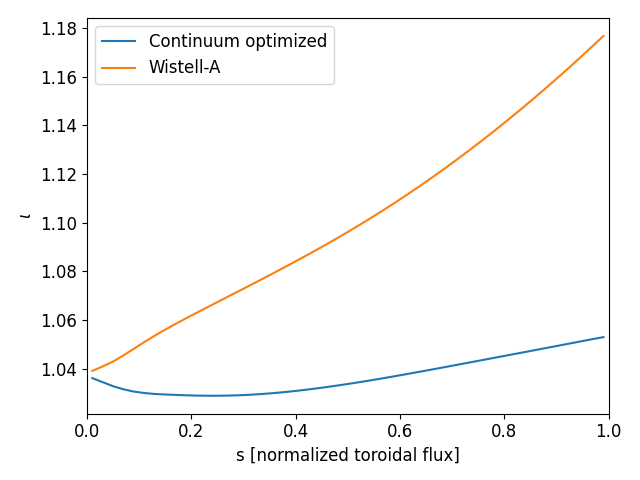}
\caption{}
    \end{subfigure}
    \begin{subfigure}{0.49\textwidth}
\includegraphics[width=0.99\textwidth]{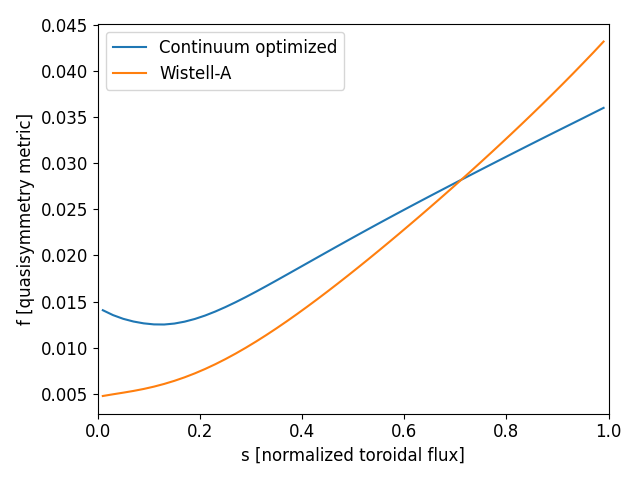}
\caption{}
    \end{subfigure}
    \begin{subfigure}{0.49\textwidth}
\includegraphics[width=0.99\textwidth]{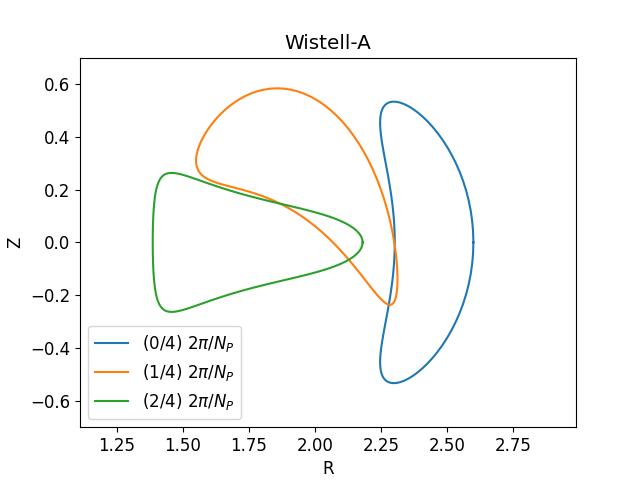}
\caption{}
    \end{subfigure}
    \begin{subfigure}{0.49\textwidth}
\includegraphics[width=0.99\textwidth]{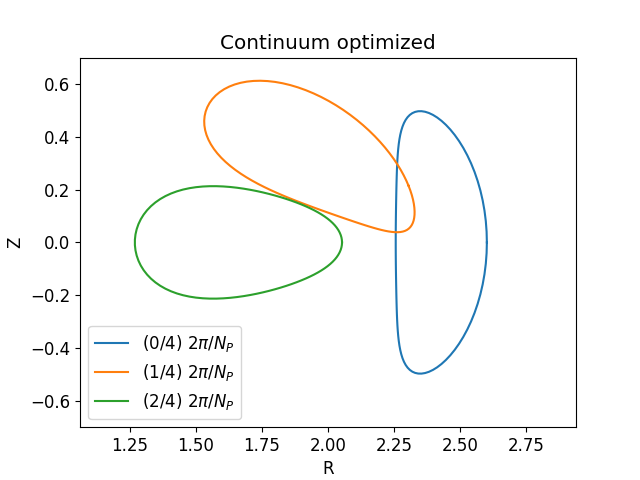}
\caption{}
    \end{subfigure}
        \begin{subfigure}{0.49\textwidth}
\includegraphics[width=0.99\textwidth]{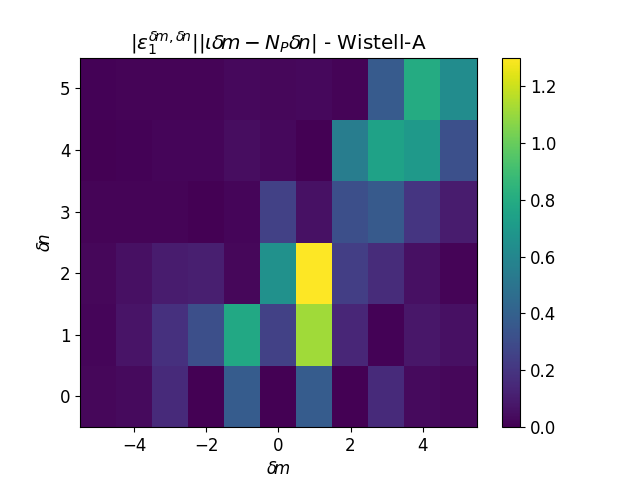}
\caption{}
    \end{subfigure}
    \begin{subfigure}{0.49\textwidth}
\includegraphics[width=0.99\textwidth]{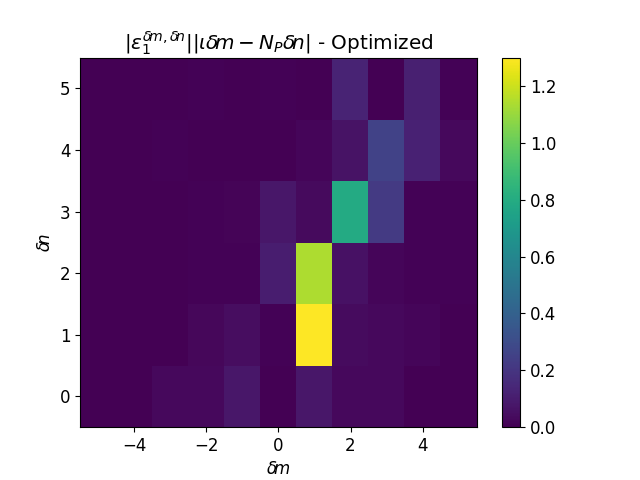}
\caption{}
    \end{subfigure}
    \caption{The rotational transform profiles (a), quasisymmetry error (b), boundary shapes (c) and (d), and spectral content of $|\nabla \psi|^2$ (e) and (f) are compared for the Wistell-A and continuum optimized configurations.}
    \label{fig:continuum_opt}
\end{figure}

At each function evaluation, a fixed-boundary VMEC is computed with the prescribed plasma boundary. A Boozer transformation is performed with booz\_xform \citep{2000Sanchez}, and a Fourier transform of $\epsilon_1$ \eqref{eq:def_eps_1_2} is computed analogously to \eqref{eq:Fourier_epsilon}. 
The optimized configuration obtained with SIMSOPT \citep{2021Landreman} is compared with the Wistell-A configuration in Figure \ref{fig:continuum_opt}. We note that the level of quasisymmetry error is roughly maintained, while the shear in the rotational transform is significantly reduced. The spectral content of $|\nabla \psi|^2$ indicates significant contributions of large mode number components to the gap width, due to the scaling with the frequency factor, $|\iota \deltam - N_P \deltan|$. In the Wistell-A configuration, significant mirror and elliptical modes are present in addition to many helical modes. 
In the optimized configuration, the spectral content is markedly reduced, primarily represented by helical (3,2), (2,1), and (1,1) components. The boundary shape becomes visually more elliptical. 

\begin{figure}
    \centering
    \begin{subfigure}{0.49\textwidth}
    \centering\includegraphics[width=1.11\linewidth]{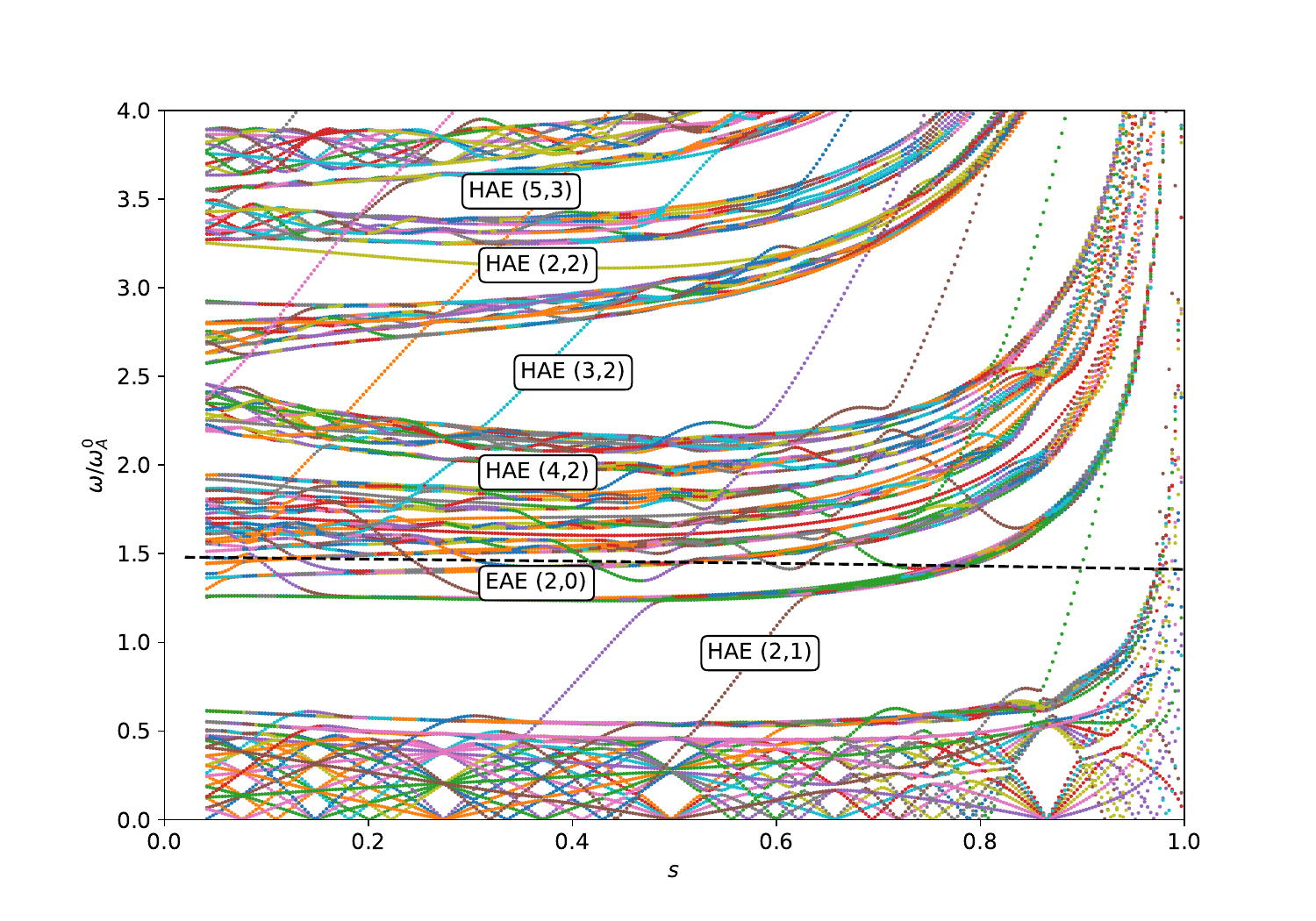}
    \caption{}
    \end{subfigure}
    \begin{subfigure}{0.49\textwidth} 
    \centering \includegraphics[width=1.12\linewidth]{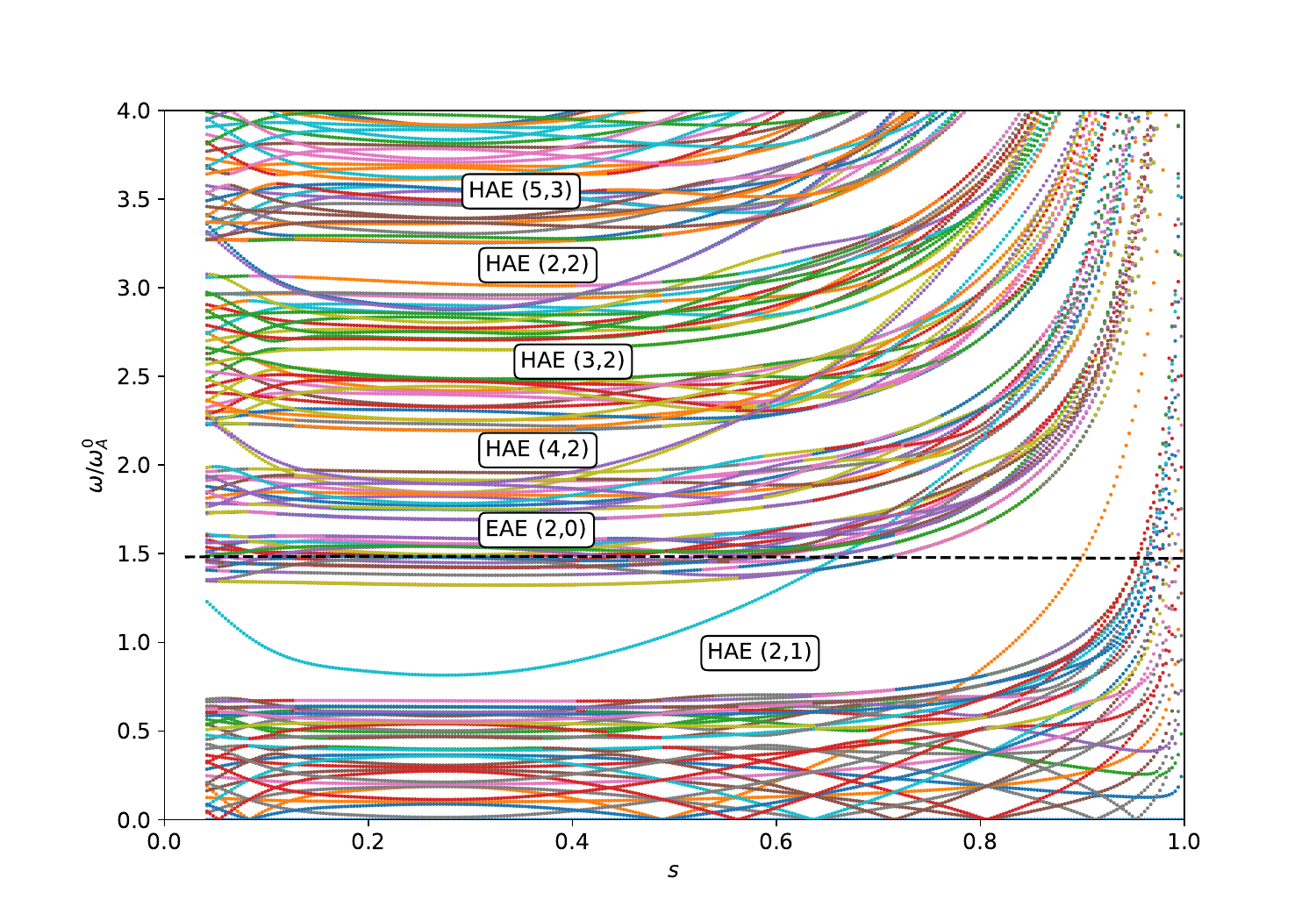}
    \caption{}
    \end{subfigure}
    \caption{The shear Alfv\'{e}n continuum is computed for the Wistell-A (left) and continuum optimized configuration (right) showing a significant reduction in high-frequency gap widths (above $\omega/\omega_A^0 = |\iota - N|/2$, indicated by horizontal dashed line). }
    \label{fig:aten_opt_continuum}
\end{figure}

The continuum of the Wistell-A and optimized configurations are computed with STELLGAP \citep{2003Spong}, shown in Figure \ref{fig:aten_opt_continuum}. The configurations are scaled to ARIES-CS volume and field strength \citep{2008Najmabadi}, with density profiles roughly consistent with ARIES-CS \citep{2020Bader}. Given the density profile shear, the eigenmode frequencies are normalized by the Alfv\'{e}n frequency on the magnetic axis, $\omega_A^0$. We see a significant reduction in the high-frequency gap widths, especially the HAE (3,2). Here the objective function is penalizing gaps above $\omega/\omega_A^0 = |\iota - N|/2 \approx 1.5$. A few higher-frequency gaps remain, such as the HAE (4,2) and (2,2), which are formed due to the nonlinear interaction of the (1,1) and (2,1) harmonics of $|\nabla \psi|^2$ not accounted for by our optimization metric. Future work will further refine this optimization strategy to account for non-perturbative impact of the geometry on gap formation, such as through direction calculation of the shear Alfv\'{e}n continuum with spectral density methods \citep{2006Weisse}.

\section{Conclusions}

The structure of the shear Alfv\'{e}n continuum is driven by the flux-surface compression factor $|\nabla \psi|^2$ and can play a critical role in the determination of stability to energetic particle driven modes. We have analyzed the impact of quasisymmetric geometry on the continuum. A near-axis model is used to determine the dominant spectral components of $|\nabla \psi|^2$. The rotating elliptical flux-surface shapes near the axis provide a helical $m = 2$, $n = N_P$ structure, which is associated with a helicity-induced Alfv\'{e}n eigenmode (HAE) gap. The toroidal variation of the ellipticity and axis curvature is shown to give rise to mirror Alfv\'{e}n eigenmode (MAE) gaps, while elongation is shown to drive ellipticity-induced Alfv\'{e}n eigenmodes (EAE) gaps in QA configurations and $m =2$, $n = 2 N_P$ gaps in QH configurations. These observations are shown to be consistent with a set of numerically-optimized quasisymmetric configurations. Through perturbative analysis, we highlight features of continuum solutions in quasisymmetric geometry, such as the presence of co-propagating continuum modes and high-order crossings. Both of these features are unique to 3D systems and lead to continuum modes which appear to cross the spectral gaps. In a follow-up paper, we will analyze the continuum of several optimized QS configurations and compare with the predictions from this theory.  

Because of the connection between continuum gaps and flux-surface shaping, we describe one strategy to manipulate the geometry to favorably modify the gap structure for EP stability of quasisymmetric stellarator reactors. Namely, by promoting wide gaps at low frequency at which passing resonant interactions are less likely to occur, higher-frequency gaps will be narrowed. This will, in turn, increase continuum damping of AEs which can strongly resonate with alpha particles near the birth energy in stellarator reactors. We demonstrate our optimization technique, producing a QH configuration with reduced high-frequency gap widths. In this case, the remaining low-frequency HAE gap did not increase substantially in width. Thus, overall, we anticipate enhanced continuum damping in the optimized configuration. Further analysis of this configuration is necessary to demonstrate an impact on the growth rates of gap AEs. Future work will refine this criterion to also account for trapped particle resonances and other passing resonances that may arise as alphas slow down. 
We furthermore, remark that the simplified metric presented assumes an analytic model for gap width. This assumption could be relaxed by directly computing the continuum within the optimization loop, using the spectral density \citep{2006Weisse} as an optimization target. Beyond stellarator design, 3D tokamak control coils could be optimized to favorably modify the continuum structure for control of EP instabilities \citep{2019Garcia}.  

 There may be other applications for modification of the in-surface variation of $|\nabla \psi|^2$. For example, modulation of this quantity can provide flow-shear stabilization \citep{Spong2007}, modify the drive for microinstabilities \citep{2024Roberg}, and affect the zonal flow residual \citep{2025Zhu, rodriguez2024zonal}. 

\section*{Acknowledgements}

The authors would like to acknowledge discussions with D. Spong and A. K\"{o}nies. 

\section*{Supplementary material}

Supplementary material is available at \href{https://doi.org/10.5281/zenodo.15520685}{https://doi.org/10.5281/zenodo.15520685}.

\section*{Funding}

We acknowledge funding through the U. S. Department of Energy, under contracts DE-SC0024630, DE-SC0024548, and DE-AC02-09CH11466. We also acknowledge funding through the Simons Foundation collaboration ``Hidden Symmetries and Fusion Energy,'' Grant No. 601958. This research used resources of the National Energy Research Scientific Computing Center (NERSC), a Department of Energy Office of Science User Facility using NERSC award ERCAP0031926. E. R. was supported by a grant by Alexander-von-Humboldt-Stiftung, Bonn, Germany, through a postdoctoral research fellowship. R. J was supported by the National Science Foundation under Grant No. 2409066.

\section*{Declaration of interests}

The authors report no conflict of interest.

\appendix
\section{Near-axis ellipse properties}
\label{app:NAE_ellipse}

The position vector in the near-axis quasisymmetry model reads \citep{2018Landreman}:
\begin{align}
    \bm{r}(r,\chi,\zeta) = \bm{r}_0(\zeta) + \frac{r}{\overline{\kappa}} \left[\cos \chi \hat{\bm{n}} + \overline{\kappa}^2 \left(\sin \chi + \sigma \cos \chi \right)\hat{\bm{b}}\right].
    \label{eq:nae_position}
\end{align}
We define the coordinates $X = \cos \chi/\overline{\kappa}$ and $Y = \overline{\kappa}(\sin \chi + \sigma \cos \chi)$ which span the perpendicular plane and the coordinates $u$ and $v$ that span the semi-major and minor axes of the elliptical flux surfaces. The angle $\vartheta$ parameterizes the ellipse such that $u = a \cos \vartheta$ and $v = b \sin \vartheta$, where $a$ is the semi-major axis and $b$ is the semi-minor axis. Following \cite{2023Rodriguez}, a quadratic equation for $X$ and $Y$ can obtained from the condition $\cos^2 \chi + \sin^2 \chi = 1$, 
\begin{align}
    X^2 \overline{\kappa}^2 (1 + \sigma^2) + \frac{Y}{\overline{\kappa}^2} - 2 \sigma XY = 1. 
    \label{eq:general_ellipse}
\end{align}
This is in the form of a general ellipse, which can be solved for the semi-major and semi-minor axes as:
\begin{align}
      a^2,b^2   &= \frac{p \pm \sqrt{p^2-4}}{2},
\end{align}
where $p = (\overline{\kappa}^4(1+\sigma^2) + 1)/\overline{\kappa}^2$. The elongation $\mathcal{E}$ is the ratio of the axes, as shown in \eqref{eq:elongation}. The general form of the ellipse \eqref{eq:general_ellipse} also provides the rotation angle $\gamma$ between the ellipse major axis and the normal vector, as shown in \eqref{eq:ellipse_rotation}.

The position vector can now be written in the ellipse coordinate system $(r,\vartheta,\zeta)$ as:
\begin{align}
    \bm{r}(r,\vartheta,\zeta) = \bm{r}_0(\zeta) + r \left[a(\zeta) \cos(\vartheta) \hat{\bm{x}}(\zeta) + b(\zeta) \sin(\vartheta) \hat{\bm{y}}(\zeta)\right]. 
\end{align}
The quantity $\nabla \psi = \left(\partial \bm{r}/\partial \vartheta \times \partial \bm{r}/\partial \zeta\right)/\left(\partial \bm{r}/\partial \psi \times \partial \bm{r}/\partial \vartheta \cdot \partial \bm{r}/\partial \zeta \right)$ can then be evaluated from derivatives of the position vector, resulting in the expression \eqref{eq:gradr2_ellipse}.

The rotation between the normal vector and the $\chi$ contours is also determined from \eqref{eq:nae_position}. Following \cite{2018Landreman}, we can consider, for example, the rotation between $\hat{\bm{n}}$ and the $\chi = 0$ contour. The vector pointing from the magnetic axis to the position at $(r,\chi=0,\zeta)$ is:
\begin{align}
    \partder{\bm{r}(r, \chi=0, \zeta)}{r} = \frac{1}{\overline{\kappa}(\zeta)} \left[\hat{\bm{n}}(\zeta) + \overline{\kappa}(\zeta)^2\sigma(\zeta) \hat{\bm{b}}(\zeta)\right].
\end{align}
The angle $\delta(\zeta)$ between $\partial \bm{r}/\partial r \rvert_{\chi = 0}$ and $\hat{\bm{n}}$ then satisfies 
\begin{align}
    \cos(\delta) = \frac{1}{\overline{\kappa}\sqrt{p-\overline{\kappa}^2}}. 
    \label{eq:cosdelta}
\end{align}
Furthermore, quasisymmetry constrains the angle between the ellipse semi-major axis and the normal vector, $\gamma(\zeta)$ \citep{2023Rodriguez}, 
\begin{align}
  \tan(2 \gamma) =
  \frac{2\sqrt{p \overline{\kappa}^2 - 1 - \overline{\kappa}^4}}{2 - p \overline{\kappa}^2}.
    \label{eq:ellipse_rotation}
\end{align}
We conclude that the rotation between the ellipse parameterization angle $\vartheta$ and the Boozer angles arises due to $\delta$ and $\gamma$ (which depend on $p$ and $\overline{\kappa}$). 



\section{Second-order degenerate perturbation theory}
\label{app:second_order_degeneracy}

This Appendix reviews details of the second-order degenerate perturbation theory presented in Section \ref{sec:higher_degeneracy}. We integrate the second-order equation \eqref{eq:2nd_order_degenerate_continuum}, repeated here for convenience,
\begin{multline}
    \left(\partder{}{\zeta} + \iota_0 \partder{}{\theta} \right)^2 \Phi^{(2)} +  
    \left[\left(\partder{}{\zeta} + \iota_0 \partder{}{\theta} \right) \epsilon \right]\left[\left(\partder{}{\zeta} + \iota_0 \partder{}{\theta} \right) \Phi^{(1)}  \right] \\
    - \epsilon \left[\left(\partder{}{\zeta} + \iota_0 \partder{}{\theta} \right) \epsilon \right]\left[\left(\partder{}{\zeta} + \iota_0 \partder{}{\theta} \right) \Phi^{(0)}  \right]  \\
  + \lambda^{(0)}  \Phi^{(2)} + \lambda^{(2)} \Phi^{(0)}  = 0,
    \label{eq:2nd_order_degenerate_continuum_app}
\end{multline}
against $\left(\Phi_j^{(0)}\right)^*$ and $\left(\Phi_k^{(0)}\right)^*$. In doing so, integrating against the first term will cancel with the integral against the fourth term. Integration of the second term against $\left(\Phi_j^{(0)}\right)^*$ can be expressed in the following form:
\begin{align}
   \frac{1}{4\pi^2}\int_0^{2\pi} d\theta \int_0^{2\pi} d \zeta \, \left(\Phi_j^{(0)}\right)^*\left[\left(\partder{}{\zeta} + \iota_0 \partder{}{\theta} \right) \epsilon \right]\left[\left(\partder{}{\zeta} + \iota_0 \partder{}{\theta} \right) \Phi^{(1)}  \right] = A_{jj} \alpha_j + A_{jk} \alpha_k,
\end{align}
with,
\begin{align}
A_{jj} &= \sum_i\Delta_{ij}^2|\epsilon_{ij}|^2 \frac{\overline{\omega}_i^{(0)}\overline{\omega}_j^{(0)}}{\lambda^{(0)}-\lambda_i^{(0)}}, \\
A_{jk} &= \sum_i \epsilon_{ij}^*\epsilon_{ik}\frac{\overline{\omega}_i^{(0)} \lambda^{(0)}}{\lambda^{(0)}-\lambda_i^{(0)}},
\end{align}
$\Delta_{ij} = \overline{\omega}_i^{(0)}-\overline{\omega}_j^{(0)}$, $\epsilon_{ij} = \epsilon_{m_i-m_j,n_i-n_j}$, and $\lambda_i^{(0)} = \left(\overline{\omega}_i^{(0)}\right)^2$. 

We now simplify using the assumption of counter-propagation. We define the index variables $\deltam = m_i-m_j$ and $\deltan = (n_i-n_j)/N_P$ such that $\epsilon_{\deltam,\deltan}=\epsilon_{ij}$. We also define the notation $\Delta({\deltam,\deltan}) = \Delta_{ij} = \iota_0 \deltam-N_P\deltan$, resulting in the expressions:
\begin{align}
    A_{jj} &= \lambda^{(0)} \sum_{\deltam,\deltan}  |\epsilon_{\deltam,\deltan}|^2 \frac{\Delta({\deltam,\deltan})}{\Delta({\deltam,\deltan}) + 2 \overline{\omega}^{(0)}_j}, \\
   A_{jk} &= \overline{\omega}_j^{(0)}\sum_{\deltam,\deltan} \epsilon_{\deltam,\deltan} \epsilon_{\Delta m-\deltam,\frac{\Delta n}{N_P}-\deltan} \left(\overline{\omega}^{(0)}_j - \Delta({\deltam,\deltan})\right).
\end{align}
$A_{jk}$ can be shown to vanish by manipulating the following sum using Parseval's theorem and integration by parts:
\begin{align*}
   \sum_{\deltam,\deltan}\epsilon_{\deltam,\deltan} \epsilon_{\Delta m-\deltam,\frac{\Delta n}{N_P}-\deltan}\Delta({\deltam,\deltan}) 
    &= \frac{1}{2i} \frac{1}{4\pi^2}\int_0^{2\pi}\int_0^{2\pi} d\theta d \zeta \, e^{-i(\Delta m \theta -\Delta n\zeta)} \left(\partder{}{\zeta} + \iota_0 \partder{}{\theta} \right) |\epsilon|^2  \\
    &= \overline{\omega}_j^{(0)} \frac{1}{4\pi^2}\int_0^{2\pi}\int_0^{2\pi} \, |\epsilon|^2 e^{-i(\Delta m \theta -\Delta n\zeta)} \\
    &= \overline{\omega}_j^{(0)} \sum_{\deltam,\deltan}\epsilon_{\deltam,\deltan} \epsilon_{\Delta m-\deltam,\frac{\Delta n}{N_P}-\deltan}.
\end{align*}
$A_{kk}$ and $A_{kj}$ are similarly defined by integrating against $\left(\Phi_k^{(0)}\right)^*$, obtaining $A_{kk} = A_{jj}$ and $A_{kj} = A_{jk} = 0$. 

The integral of $\left(\Phi_j^{(0)}\right)^*$ against the third term can be written in the form:
\begin{align}
   -\frac{1}{4\pi^2}\int_0^{2\pi} d\theta \int_0^{2\pi} d \zeta \, \left(\Phi_j^{(0)}\right)^* \epsilon \left[\left(\partder{}{\zeta} + \iota_0 \partder{}{\theta} \right) \epsilon \right]\left[\left(\partder{}{\zeta} + \iota_0 \partder{}{\theta} \right) \Phi^{(0)}  \right] = B_{jk} \alpha_k
\end{align}
with 
\begin{align}
    B_{jk} = -\lambda^{(0)} \sum_{\deltam,\deltan} \epsilon_{\delta m,\delta n}  \epsilon_{\Delta m-\deltam,\Delta n/N_P-\deltan}.
\end{align}
$B_{kj}$ is defined similarly by integrating against $\left(\Phi_k^{(0)}\right)^*$, yielding $B_{kj}=B_{jk}^*$.
The zero-determinant condition then reads:
\begin{align}
        \left(\lambda^{(2)} + A_{jj}\right)^2 - |B_{jk}|^2 = 0.
\end{align}
Evidently, both the central frequency and gap width are modified at this order, with expressions given by $\sqrt{\lambda^{(0)}-A_{jj}}$ and 
\begin{align}
    \Delta \overline{\omega} = \frac{|B_{jk}|}{\overline{\omega}^{(0)}},
\end{align}
respectively.

\section{Third-order near-axis geometric factor $|\nabla \psi|^2$}
\label{app:third_order_nablapsi2}

In this Appendix, we derive the expression for the third-order $\Psi_3$ geometric factor of $|\nabla \psi|^2$ from \eqref{eq:psi_expansion}.
Due to the additional geometric effects such as Shafranov shift and triangularity, at this order, the near-axis expansion yields a total of nine new functions of $\zeta$, namely $X_{20}, X_{2c}, X_{2s}, Y_{20}, Y_{2c}, Y_{2s}, Z_{20}, Z_{2c}$ and $Z_{2s}$.
Their corresponding equations can be found in \cite{2019Landremanb}.
Also, three new input scalars ($p_2, B_{2c}, B_{2s}$) appear, where $p_2$ describes the pressure gradient and  $B_{2c}$, $B_{2s}$ describe the magnetic field strength $B$ at second order. The parameter $B_{20}$ is then determined from the provided inputs. While $B_{20}$ is a constant scalar in perfect QS, it is more generally a function of $\zeta$. 

We use the dual relations to write the geometric factor as
\begin{equation}
    \nabla r = \frac{1}{\sqrt{g}}\frac{\partial \bm{r}}{\partial \chi}\times \frac{\partial \bm{r}}{\partial \zeta},
\end{equation}
where $\sqrt{g} = \left(\partial \bm{r}/\partial r \times \partial \bm{r}/\partial \chi\right)\cdot \partial \bm{r}/\partial \zeta$ is the Jacobian, $\bm{r}$ is the position vector described using the near-axis decomposition $\bm r = \bm r_0 + X \hat{\bm{n}} + Y \hat{\bm{b}} + Z \hat{\bm{t}}$, and $(\hat{\bm{n}}, \hat{\bm{b}}, \hat{\bm{t}})$ is the Frenet-Serret frame \citep{2020Jorgeb}.
This leads to the following expression for the third-order geometric factor
%
%
%
\begin{equation}
    \Psi_3 = \Psi_{31c} \cos \chi + \Psi_{31s} \sin \chi + \Psi_{33c} \cos 3 \chi + \Psi_{33s} \sin 3 \chi,
\end{equation}
where
\begin{multline}
    \Psi_{31c} = \frac{B_{0}^2}{\overline{\kappa}^3} \Big(-Y_{22s}+ 
    \overline{\kappa}^2 (-X_{20}+X_{22c}+\sigma X_{22s}) \\ 
    - \overline{\kappa}^4 \left(\left(\sigma^2+1\right) Y_{22s}-2 \sigma Y_{20}\right)\\
    +\overline{\kappa}^6 \left(-3 \left(\sigma^2+1\right) X_{20}+\left(\sigma^2-3\right) X_{22c}+\sigma \left(\sigma^2+5\right) X_{22s}\right)\Big),
\end{multline}
\begin{multline}
    \Psi_{31s} = \frac{B_{0}^2}{\overline{\kappa}^3} \Big(3 (Y_{22c}-Y_{20})-\overline{\kappa}^2 (3 \sigma (X_{22c}-X_{20})+X_{22s})\\
    -\overline{\kappa}^4 \left(Y_{20}+3 \sigma^2 (Y_{20}-Y_{22c})+Y_{22c}-4 \sigma Y_{22s}\right) \\
    +\overline{\kappa}^6 \left(\sigma \left(3 X_{20}+3 \sigma^2 (X_{20}-X_{22c})+X_{22c}\right)-\left(5 \sigma^2+1\right) X_{22s}\right)\Big),
\end{multline}
\begin{multline}
    \Psi_{33c} = \frac{B_{0}^2}{\overline{\kappa}^3} \Big(Y_{22s} + \overline{\kappa}^2 (X_{20}-X_{22c}-\sigma X_{22s})\\
    +\overline{\kappa}^4 \left(\left(\sigma^2+1\right) Y_{22s}-2 \sigma Y_{20}\right) \\
    +\overline{\kappa}^6 \left(\left(3 \sigma (\Phi )^2-1\right) X_{20}-\left(\sigma^2+1\right) (X_{22c}+\sigma X_{22s})\right)\Big),
\end{multline}
\begin{multline}
    \Psi_{33s} = \frac{B_{0}^2 }{\overline{\kappa}^3}\Big((Y_{20}-Y_{22c}) -\overline{\kappa}^2 (\sigma (X_{20}-X_{22c})+X_{22s})\\
    + \overline{\kappa}^4 \left(\left(\sigma^2-1\right) Y_{20}-\left(\sigma^2+1\right) Y_{22c}\right) \\
    -\overline{\kappa}^6 \left(\sigma^3 (X_{20}-X_{22c})-\sigma (3 X_{20}+X_{22c})+\left(\sigma^2+1\right) X_{22s}\right)\Big).
\end{multline}

\bibliographystyle{jpp}

\bibliography{jpp-instructions}

\begin{thebibliography}{62}
\expandafter\ifx\csname natexlab\endcsname\relax\def\natexlab#1{#1}\fi
\def\au#1{#1} \def\ed#1{#1} \def\yr#1{#1}\def\at#1{#1}\def\jt#1{\textit{#1}} \def\bt#1{#1}\def\bvol#1{\textbf{#1}} \def\vol#1{#1} \def\pg#1{#1} \def\publ#1{#1}\def\arxiv#1{#1}\def\org#1{#1}\def\st#1{\textit{#1}}

\bibitem[Anderson {\em et~al.\/}(1995)Anderson, Almagri, Anderson, Matthews, Talmadge \& Shohet]{1995Anderson}
{\sc \au{Anderson, F. S.~B.}, \au{Almagri, A.~F.}, \au{Anderson, D.~T.}, \au{Matthews, P.~G.}, \au{Talmadge, J.~N.} \& \au{Shohet, J.~L.}} \yr{1995}  \at{{The Helically Symmetric Experiment (HSX) goals, design and status}}.  \jt{Fusion Technology}  \bvol{27}~(3T),  \pg{273--277}.

\bibitem[Bader {\em et~al.\/}(2020)Bader, Faber, Schmitt, Anderson, Drevlak, Duff, Frerichs, Hegna, Kruger, Landreman {\em et~al.\/}]{2020Bader}
{\sc \au{Bader, A.}, \au{Faber, B.~J.}, \au{Schmitt, J.~C.}, \au{Anderson, D.~T.}, \au{Drevlak, M.}, \au{Duff, J.~M.}, \au{Frerichs, H.}, \au{Hegna, C.~C.}, \au{Kruger, T.~G.}, \au{Landreman, M.} \& \au{others}} \yr{2020}  \at{Advancing the physics basis for quasi-helically symmetric stellarators}.  \jt{Journal of Plasma Physics}  \bvol{86}~(5),  \pg{905860506}.

\bibitem[Beidler {\em et~al.\/}(2001{\natexlab{{\em a\/}}})Beidler, Harmeyer, Herrnegger, Igitkhanov, Kendl, Kisslinger, Kolesnichenko, Lutsenko, N{\"u}hrenberg, Sidorenko, Strumberger, Wobig \& Yakovenko]{2001Beidler}
{\sc \au{Beidler, C.~D.}, \au{Harmeyer, E.}, \au{Herrnegger, F.}, \au{Igitkhanov, Yu.}, \au{Kendl, A.}, \au{Kisslinger, J.}, \au{Kolesnichenko, Ya.~I.}, \au{Lutsenko, V.~V.}, \au{N{\"u}hrenberg, C.}, \au{Sidorenko, I.}, \au{Strumberger, E.}, \au{Wobig, H.} \& \au{Yakovenko, Yu.~V.}} \yr{2001{\natexlab{{\em a\/}}}}  \at{The {{Helias}} reactor {{HSR4}}/18}.  \jt{Nuclear Fusion}  \bvol{41}~(12),  \pg{1759--1766}.

\bibitem[Beidler {\em et~al.\/}(2001{\natexlab{{\em b\/}}})Beidler, Kolesnichenko, Marchenko, Sidorenko \& Wobig]{2001Beidlerb}
{\sc \au{Beidler, C.~D.}, \au{Kolesnichenko, Ya~I.}, \au{Marchenko, V.~S.}, \au{Sidorenko, I.~N.} \& \au{Wobig, H.}} \yr{2001{\natexlab{{\em b\/}}}}  \at{Stochastic diffusion of energetic ions in optimized stellarators}.  \jt{Physics of Plasmas}  \bvol{8}~(6),  \pg{2731--2738}.

\bibitem[Betti \& Freidberg(1991)]{Betti1991}
{\sc \au{Betti, R.} \& \au{Freidberg, J.~P.}} \yr{1991}  \at{Ellipticity induced {{Alfv{\'e}n}} eigenmodes}.  \jt{Physics of Fluids B: Plasma Physics}  \bvol{3}~(8),  \pg{1865--1870}.

\bibitem[Boozer(1981)]{1981Boozer}
{\sc \au{Boozer, A.~H.}} \yr{1981}  \at{Plasma equilibrium with rational magnetic surfaces}.  \jt{Physics of Fluids}  \bvol{24},  \pg{1999}.

\bibitem[Boozer(1983)]{1983Boozer}
{\sc \au{Boozer, A.~H.}} \yr{1983}  \at{Transport and isomorphic equilibria}.  \jt{Physics of Fluids}  \bvol{26}~(2),  \pg{496--499}.

\bibitem[Chen \& Zonca(1995)]{1995Chen}
{\sc \au{Chen, L.} \& \au{Zonca, F.}} \yr{1995}  \at{Theory of shear {{Alfv{\'e}n}} waves in toroidal plasmas}.  \jt{Physica Scripta}  \bvol{T60},  \pg{81--90}.

\bibitem[Cheng \& Chance(1986)]{1986Cheng}
{\sc \au{Cheng, C.~Z.} \& \au{Chance, M.~S.}} \yr{1986}  \at{{Low‐n shear Alfvén spectra in axisymmetric toroidal plasmas}}.  \jt{The Physics of Fluids}  \bvol{29}~(11),  \pg{3695--3701}.

\bibitem[Fesenyuk {\em et~al.\/}(2004)Fesenyuk, Kolesnichenko, Lutsenko, White \& Yakovenko]{2004Fesenyuk}
{\sc \au{Fesenyuk, O.~P.}, \au{Kolesnichenko, Ya~I.}, \au{Lutsenko, V.~V.}, \au{White, R.~B.} \& \au{Yakovenko, Yu~V.}} \yr{2004}  \at{{Alfv{\'e}n continuum and Alfv{\'e}n eigenmodes in the National Compact Stellarator Experiment}}.  \jt{Physics of Plasmas}  \bvol{11}~(12),  \pg{5444--5451}.

\bibitem[Fesenyuk {\em et~al.\/}(2002)Fesenyuk, Kolesnichenko, Wobig \& Yakovenko]{Fesenyuk2002}
{\sc \au{Fesenyuk, O.~P.}, \au{Kolesnichenko, {\relax Ya}.~I.}, \au{Wobig, H.} \& \au{Yakovenko, {\relax Yu}.~V.}} \yr{2002}  \at{Ideal magnetohydrodynamic equations for low-frequency waves in toroidal plasmas}.  \jt{Physics of Plasmas}  \bvol{9}~(5),  \pg{1589--1595}.

\bibitem[Garcia-Munoz {\em et~al.\/}(2019)Garcia-Munoz, Sharapov, Van~Zeeland, Ascasibar, Cappa, Chen, Ferreira, Galdon-Quiroga, Geiger, Gonzalez-Martin {\em et~al.\/}]{2019Garcia}
{\sc \au{Garcia-Munoz, M.}, \au{Sharapov, S.~E.}, \au{Van~Zeeland, M.~A.}, \au{Ascasibar, E.}, \au{Cappa, A.}, \au{Chen, L.}, \au{Ferreira, J.}, \au{Galdon-Quiroga, J.}, \au{Geiger, B.}, \au{Gonzalez-Martin, J.} \& \au{others}} \yr{2019}  \at{{Active control of Alfv{\'e}n eigenmodes in magnetically confined toroidal plasmas}}.  \jt{Plasma Physics and Controlled Fusion}  \bvol{61}~(5),  \pg{054007}.

\bibitem[Garren \& Boozer(1991)]{1991Garren}
{\sc \au{Garren, D.~A.} \& \au{Boozer, A.~H.}} \yr{1991}  \at{Magnetic field strength of toroidal plasma equilibria}.  \jt{Physics of Fluids B: Plasma Physics}  \bvol{3}~(10),  \pg{2805--2821}.

\bibitem[Gorelenkov {\em et~al.\/}(2014)Gorelenkov, Pinches \& Toi]{2014Gorelenkov}
{\sc \au{Gorelenkov, N.~N.}, \au{Pinches, S.~D.} \& \au{Toi, K.}} \yr{2014}  \at{Energetic particle physics in fusion research in preparation for burning plasma experiments}.  \jt{Nuclear Fusion}  \bvol{54}~(12),  \pg{125001}.

\bibitem[Grieger {\em et~al.\/}(1992)Grieger, Lotz, Merkel, N{\"u}hrenberg, Sapper, Strumberger, Wobig, Burhenn, Erckmann, Gasparino {\em et~al.\/}]{1992Grieger}
{\sc \au{Grieger, G.}, \au{Lotz, W.}, \au{Merkel, P.}, \au{N{\"u}hrenberg, J.}, \au{Sapper, J.}, \au{Strumberger, E.}, \au{Wobig, H.}, \au{Burhenn, R.}, \au{Erckmann, V.}, \au{Gasparino, U.} \& \au{others}} \yr{1992}  \at{Physics optimization of stellarators}.  \jt{Physics of Fluids}  \bvol{4}~(7),  \pg{2081--2091}.

\bibitem[Heidbrink(2008)]{2008Heidbrink}
{\sc \au{Heidbrink, W.~W.}} \yr{2008}  \at{Basic physics of {{Alfv{\'e}n}} instabilities driven by energetic particles in toroidally confined plasmas}.  \jt{Physics of Plasmas}  \bvol{15}~(5),  \pg{055501}.

\bibitem[Jorge \& Landreman(2020)]{2020Jorge}
{\sc \au{Jorge, R.} \& \au{Landreman, M.}} \yr{2020}  \at{The use of near-axis magnetic fields for stellarator turbulence simulations}.  \jt{Plasma Physics and Controlled Fusion}  \bvol{63}~(1),  \pg{014001}.

\bibitem[Jorge {\em et~al.\/}(2020{\natexlab{{\em a\/}}})Jorge, Sengupta \& Landreman]{2020Jorgec}
{\sc \au{Jorge, R.}, \au{Sengupta, W.} \& \au{Landreman, M.}} \yr{2020{\natexlab{{\em a\/}}}}  \at{Construction of quasisymmetric stellarators using a direct coordinate approach}.  \jt{Nuclear Fusion}  \bvol{60}~(7),  \pg{076021}.

\bibitem[Jorge {\em et~al.\/}(2020{\natexlab{{\em b\/}}})Jorge, Sengupta \& Landreman]{2020Jorgeb}
{\sc \au{Jorge, R.}, \au{Sengupta, W.} \& \au{Landreman, M.}} \yr{2020{\natexlab{{\em b\/}}}}  \at{Near-axis expansion of stellarator equilibrium at arbitrary order in the distance to the axis}.  \jt{Journal of Plasma Physics}  \bvol{86}~(1),  \pg{905860106}.

\bibitem[Kieras \& Tataronis(1982)]{1982Kieras}
{\sc \au{Kieras, C.~E.} \& \au{Tataronis, J.~A.}} \yr{1982}  \at{The shear {Alfv{\'e}n} continuous spectrum of axisymmetric toroidal equilibria in the large aspect ratio limit}.  \jt{Journal of Plasma Physics}  \bvol{28}~(3),  \pg{395--414}.

\bibitem[Kolesnichenko {\em et~al.\/}(2011)Kolesnichenko, K{\"o}nies, Lutsenko \& Yakovenko]{2011Kolesnichenko}
{\sc \au{Kolesnichenko, Y.~I.}, \au{K{\"o}nies, A.}, \au{Lutsenko, V.~V.} \& \au{Yakovenko, Y.~V.}} \yr{2011}  \at{{Affinity and difference between energetic-ion-driven instabilities in 2D and 3D toroidal systems}}.  \jt{Plasma Physics and Controlled Fusion}  \bvol{53}~(2),  \pg{024007}.

\bibitem[Kolesnichenko {\em et~al.\/}(2001)Kolesnichenko, Lutsenko, Wobig, Yakovenko \& Fesenyuk]{2001Kolesnichenko}
{\sc \au{Kolesnichenko, Y.~I.}, \au{Lutsenko, V.~V.}, \au{Wobig, H.}, \au{Yakovenko, Y.~V.} \& \au{Fesenyuk, O.~P.}} \yr{2001}  \at{{Alfv{\'e}n continuum and high-frequency eigenmodes in optimized stellarators}}.  \jt{Physics of Plasmas}  \bvol{8}~(2),  \pg{491--509}.

\bibitem[K{\"o}nies \& Eremin(2010)]{2010Konies}
{\sc \au{K{\"o}nies, A.} \& \au{Eremin, D.}} \yr{2010}  \at{{Coupling of Alfv{\'e}n and sound waves in stellarator plasmas}}.  \jt{Physics of Plasmas}  \bvol{17}~(1).

\bibitem[Kramer {\em et~al.\/}(1998)Kramer, Saigusa, Ozeki, Kusama, Kimura, Oikawa, Tobita, Fu \& Cheng]{1998Kramer}
{\sc \au{Kramer, G.~J.}, \au{Saigusa, M.}, \au{Ozeki, T.}, \au{Kusama, Y.}, \au{Kimura, H.}, \au{Oikawa, T.}, \au{Tobita, K.}, \au{Fu, G.~Y.} \& \au{Cheng, C.~Z.}} \yr{1998}  \at{Noncircular triangularity and ellipticity-induced alfv{\'e}n eigenmodes observed in jt-60u}.  \jt{Physical Review Letters}  \bvol{80}~(12),  \pg{2594}.

\bibitem[Landreman(2019)]{2019Landreman}
{\sc \au{Landreman, M.}} \yr{2019}  \at{{Optimized quasisymmetric stellarators are consistent with the Garren--Boozer construction}}.  \jt{Plasma Physics and Controlled Fusion}  \bvol{61}~(7),  \pg{075001}.

\bibitem[Landreman(2022)]{2022Landreman_mapping}
{\sc \au{Landreman, M.}} \yr{2022}  \at{Mapping the space of quasisymmetric stellarators using optimized near-axis expansion}.  \jt{Journal of Plasma Physics}  \bvol{88}~(6),  \pg{905880616}.

\bibitem[Landreman(2024)]{pyQSC}
{\sc \au{Landreman, M.}} \yr{2024} {pyQSC}. \url{https://github.com/landreman/pyQSC}, accessed: 2024-05-31.

\bibitem[Landreman {\em et~al.\/}(2022)Landreman, Buller \& Drevlak]{2022Landremanb}
{\sc \au{Landreman, M.}, \au{Buller, S.} \& \au{Drevlak, M.}} \yr{2022}  \at{Optimization of quasi-symmetric stellarators with self-consistent bootstrap current and energetic particle confinement}.  \jt{Physics of Plasmas}  \bvol{29}~(8),  \pg{082501}.

\bibitem[Landreman {\em et~al.\/}(2021)Landreman, Medasani, Wechsung, Giuliani, Jorge \& Zhu]{2021Landreman}
{\sc \au{Landreman, M.}, \au{Medasani, B.}, \au{Wechsung, F.}, \au{Giuliani, A.}, \au{Jorge, R.} \& \au{Zhu, C.}} \yr{2021}  \at{{SIMSOPT: a flexible framework for stellarator optimization}}.  \jt{Journal of Open Source Software}  \bvol{6}~(65),  \pg{3525}.

\bibitem[Landreman \& Paul(2022)]{2022Landreman}
{\sc \au{Landreman, M.} \& \au{Paul, E.}} \yr{2022}  \at{Magnetic fields with precise quasisymmetry for plasma confinement}.  \jt{Physical Review Letters}  \bvol{128}~(3),  \pg{035001}.

\bibitem[Landreman \& Sengupta(2018)]{2018Landreman}
{\sc \au{Landreman, M.} \& \au{Sengupta, W.}} \yr{2018}  \at{Direct construction of optimized stellarator shapes. {{Part}} 1. {{Theory}} in cylindrical coordinates}.  \jt{Journal of Plasma Physics}  \bvol{84}~(6),  \pg{905840616}.

\bibitem[Landreman \& Sengupta(2019)]{2019Landremanb}
{\sc \au{Landreman, M.} \& \au{Sengupta, W.}} \yr{2019}  \at{Constructing stellarators with quasisymmetry to high order}.  \jt{Journal of Plasma Physics}  \bvol{85}~(6),  \pg{815850601}.

\bibitem[Landreman {\em et~al.\/}(2019)Landreman, Sengupta \& Plunk]{2019Landremanc}
{\sc \au{Landreman, M.}, \au{Sengupta, W.} \& \au{Plunk, G.~G.}} \yr{2019}  \at{Direct construction of optimized stellarator shapes. part 2. numerical quasisymmetric solutions}.  \jt{Journal of Plasma Physics}  \bvol{85}~(1),  \pg{905850103}.

\bibitem[Mercier(1964)]{mercier1964equilibrium}
{\sc \au{Mercier, Claude}} \yr{1964}  \at{Equilibrium and stability of a toroidal magnetohydrodynamic system in the neighbourhood of a magnetic axis}.  \jt{Nuclear Fusion}  \bvol{4}~(3),  \pg{213}.

\bibitem[Mynick(2006)]{2006Mynick}
{\sc \au{Mynick, H.~E.}} \yr{2006}  \at{Transport optimization in stellarators}.  \jt{Physics of Plasmas}  \bvol{13}~(5).

\bibitem[Najmabadi {\em et~al.\/}(2008)Najmabadi, Raffray, Abdel-Khalik, Bromberg, Crosatti, El-Guebaly, Garabedian, Grossman, Henderson, Ibrahim {\em et~al.\/}]{2008Najmabadi}
{\sc \au{Najmabadi, F.}, \au{Raffray, A.~R.}, \au{Abdel-Khalik, S.~I.}, \au{Bromberg, L.}, \au{Crosatti, L.}, \au{El-Guebaly, L.}, \au{Garabedian, P.~R.}, \au{Grossman, A.~A.}, \au{Henderson, D.}, \au{Ibrahim, A.} \& \au{others}} \yr{2008}  \at{{The ARIES-CS compact stellarator fusion power plant}}.  \jt{Fusion Science and Technology}  \bvol{54}~(3),  \pg{655--672}.

\bibitem[Nies {\em et~al.\/}(2024)Nies, Paul, Panici, Hudson \& Bhattacharjee]{2024Nies}
{\sc \au{Nies, R.}, \au{Paul, E.~J.}, \au{Panici, D.}, \au{Hudson, S.~R.} \& \au{Bhattacharjee, A.}} \yr{2024}  \at{Exploration of the parameter space of quasisymmetric stellarator vacuum fields through adjoint optimisation}.  \jt{Journal of Plasma Physics}  \bvol{90}~(6),  \pg{905900620}.

\bibitem[N{\"u}hrenberg(1999)]{1999Nuhrenberg}
{\sc \au{N{\"u}hrenberg, C.}} \yr{1999}  \at{{Compressional ideal magnetohydrodynamics: Unstable global modes, stable spectra, and Alfv{\'e}n eigenmodes in Wendelstein 7--X-type equilibria}}.  \jt{Physics of Plasmas}  \bvol{6}~(1),  \pg{137--147}.

\bibitem[N{\"u}hrenberg \& Zille(1988)]{1988Nuhrenberg}
{\sc \au{N{\"u}hrenberg, J.} \& \au{Zille, R.}} \yr{1988}  \at{Quasi-helically symmetric toroidal stellarators}.  \jt{Physics Letters A}  \bvol{129}~(2),  \pg{113--117}.

\bibitem[Paul {\em et~al.\/}(2022)Paul, Bhattacharjee, Landreman, Alex, Velasco \& Nies]{2022Paul}
{\sc \au{Paul, E.~J.}, \au{Bhattacharjee, A.}, \au{Landreman, M.}, \au{Alex, D.}, \au{Velasco, J.~L.} \& \au{Nies, R.}} \yr{2022}  \at{Energetic particle loss mechanisms in reactor-scale equilibria close to quasisymmetry}.  \jt{Nuclear Fusion}  \bvol{62}~(12),  \pg{126054}.

\bibitem[Paul {\em et~al.\/}(2023)Paul, Mynick \& Bhattacharjee]{2023Paul}
{\sc \au{Paul, E.~J.}, \au{Mynick, H.~E.} \& \au{Bhattacharjee, A.}} \yr{2023}  \at{Fast-ion transport in quasisymmetric equilibria in the presence of a resonant {{Alfv{\'e}nic}} perturbation}.  \jt{Journal of Plasma Physics}  \bvol{89}~(5),  \pg{905890515}.

\bibitem[Redi {\em et~al.\/}(1999)Redi, Mynick, Suewattana, White \& Zarnstorff]{1999Fedi}
{\sc \au{Redi, M.~H.}, \au{Mynick, H.~E.}, \au{Suewattana, M.}, \au{White, R.~B.} \& \au{Zarnstorff, M.~C.}} \yr{1999}  \at{Energetic particle transport in compact quasi-axisymmetric stellarators}.  \jt{Physics of Plasmas}  \bvol{6}~(9),  \pg{3509--3520}.

\bibitem[Riyopoulos \& Mahajan(1986)]{1986Riyopoulos}
{\sc \au{Riyopoulos, S.} \& \au{Mahajan, S.}} \yr{1986}  \at{Alfv{\'e}n continuum with toroidicity}.  \jt{The Physics of Fluids}  \bvol{29}~(3),  \pg{731--740}.

\bibitem[Roberg-Clark {\em et~al.\/}(2024)Roberg-Clark, Xanthopoulos \& Plunk]{2024Roberg}
{\sc \au{Roberg-Clark, G.~T.}, \au{Xanthopoulos, P.} \& \au{Plunk, G.~G.}} \yr{2024}  \at{Reduction of electrostatic turbulence in a quasi-helically symmetric stellarator via critical gradient optimization}.  \jt{Journal of Plasma Physics}  \bvol{90}~(3),  \pg{175900301}.

\bibitem[Rodr{\'\i}guez(2023)]{2023Rodriguez}
{\sc \au{Rodr{\'\i}guez, E.}} \yr{2023}  \at{Magnetohydrodynamic stability and the effects of shaping: a near-axis view for tokamaks and quasisymmetric stellarators}.  \jt{Journal of Plasma Physics}  \bvol{89}~(2),  \pg{905890211}.

\bibitem[Rodriguez {\em et~al.\/}(2020)Rodriguez, Helander \& Bhattacharjee]{2020Rodriguez}
{\sc \au{Rodriguez, E.}, \au{Helander, P.} \& \au{Bhattacharjee, A.}} \yr{2020}  \at{Necessary and sufficient conditions for quasisymmetry}.  \jt{Physics of Plasmas}  \bvol{27}~(6).

\bibitem[Rodriguez {\em et~al.\/}(2022)Rodriguez, Paul \& Bhattacharjee]{2022Rodriguezb}
{\sc \au{Rodriguez, E.}, \au{Paul, E.~J.} \& \au{Bhattacharjee, A.}} \yr{2022}  \at{Measures of quasisymmetry for stellarators}.  \jt{Journal of Plasma Physics}  \bvol{88}~(1),  \pg{905880109}.

\bibitem[Rodriguez \& Plunk(2024)]{rodriguez2024zonal}
{\sc \au{Rodriguez, Eduardo} \& \au{Plunk, Gabriel~G}} \yr{2024}  \at{The zonal-flow residual does not tend to zero in the limit of small mirror ratio}.  \jt{arXiv preprint arXiv:2407.17824} .

\bibitem[Rodr{\'\i}guez {\em et~al.\/}(2023{\natexlab{{\em a\/}}})Rodr{\'\i}guez, Sengupta \& Bhattacharjee]{rodriguez2023constructing}
{\sc \au{Rodr{\'\i}guez, Eduardo}, \au{Sengupta, W} \& \au{Bhattacharjee, A}} \yr{2023{\natexlab{{\em a\/}}}}  \at{Constructing the space of quasisymmetric stellarators through near-axis expansion}.  \jt{Plasma Physics and Controlled Fusion}  \bvol{65}~(9),  \pg{095004}.

\bibitem[Rodr{\'\i}guez {\em et~al.\/}(2023{\natexlab{{\em b\/}}})Rodr{\'\i}guez, Sengupta \& Bhattacharjee]{2023Rodriguezb}
{\sc \au{Rodr{\'\i}guez, E.}, \au{Sengupta, W.} \& \au{Bhattacharjee, A.}} \yr{2023{\natexlab{{\em b\/}}}}  \at{Constructing the space of quasisymmetric stellarators through near-axis expansion}.  \jt{Plasma Physics and Controlled Fusion}  \bvol{65}~(9),  \pg{095004}.

\bibitem[Salat \& Tataronis(2001{\natexlab{{\em a\/}}})]{salat_shear_2001}
{\sc \au{Salat, A.} \& \au{Tataronis, J.~A.}} \yr{2001{\natexlab{{\em a\/}}}}  \at{Shear {Alfvén} mode resonances in nonaxisymmetric toroidal low-pressure plasmas. {I}. {Mode} equations in arbitrary geometry}.  \jt{Physics of Plasmas}  \bvol{8}~(4),  \pg{1200--1206}.

\bibitem[Salat \& Tataronis(2001{\natexlab{{\em b\/}}})]{salat_shear_2001-1}
{\sc \au{Salat, A.} \& \au{Tataronis, J.~A.}} \yr{2001{\natexlab{{\em b\/}}}}  \at{Shear {Alfvén} mode resonances in nonaxisymmetric toroidal low-pressure plasmas. {II}. {Singular} modes in the shear {Alfvén} continuum}.  \jt{Physics of Plasmas}  \bvol{8}~(4),  \pg{1207--1218}.

\bibitem[Sanchez {\em et~al.\/}(2000)Sanchez, Hirshman, Ware, Berry \& Spong]{2000Sanchez}
{\sc \au{Sanchez, R.}, \au{Hirshman, S.~P.}, \au{Ware, A.~S.}, \au{Berry, L.~A.} \& \au{Spong, D.~A.}} \yr{2000}  \at{Ballooning stability optimization of low-aspect-ratio stellarators}.  \jt{Plasma Physics and Controlled Fusion}  \bvol{42}~(6),  \pg{641}.

\bibitem[Spong {\em et~al.\/}(2007)Spong, Harris, Ware, Hirshman \& Berry]{Spong2007}
{\sc \au{Spong, D.~A.}, \au{Harris, J.~H.}, \au{Ware, A.~S.}, \au{Hirshman, S.~P.} \& \au{Berry, L.~A.}} \yr{2007}  \at{Shear flow generation in stellarators---configurational variations}.  \jt{Nuclear Fusion}  \bvol{47}~(7),  \pg{626--633}.

\bibitem[Spong {\em et~al.\/}(2003)Spong, Sanchez \& Weller]{2003Spong}
{\sc \au{Spong, D.~A.}, \au{Sanchez, R.} \& \au{Weller, A.}} \yr{2003}  \at{Shear {Alfv{\'e}n} continua in stellarators}.  \jt{Physics of Plasmas}  \bvol{10}~(8),  \pg{3217--3224}.

\bibitem[Toi {\em et~al.\/}(2011)Toi, Ogawa, Isobe, Osakabe, Spong \& Todo]{2011Toi}
{\sc \au{Toi, K.}, \au{Ogawa, K.}, \au{Isobe, M.}, \au{Osakabe, M.}, \au{Spong, D.~A.} \& \au{Todo, Y.}} \yr{2011}  \at{Energetic-ion-driven global instabilities in stellarator/helical plasmas and comparison with tokamak plasmas}.  \jt{Plasma Physics and Controlled Fusion}  \bvol{53}~(2),  \pg{024008}.

\bibitem[Van~Zeeland {\em et~al.\/}(2006)Van~Zeeland, Kramer, Austin, Boivin, Heidbrink, Makowski, McKee, Nazikian, Solomon \& Wang]{2006Van}
{\sc \au{Van~Zeeland, M.~A.}, \au{Kramer, G.~J.}, \au{Austin, M.~E.}, \au{Boivin, R.~L.}, \au{Heidbrink, W.~W.}, \au{Makowski, M.~A.}, \au{McKee, G.~R.}, \au{Nazikian, R.}, \au{Solomon, W.~M.} \& \au{Wang, G.}} \yr{2006}  \at{{Radial structure of Alfv{\'e}n eigenmodes in the DIII-D tokamak through electron-cyclotron-emission measurements}}.  \jt{Physical Review Letters}  \bvol{97}~(13),  \pg{135001}.

\bibitem[Varela {\em et~al.\/}(2021)Varela, Shimizu, Spong, Garcia \& Ghai]{2021Varela}
{\sc \au{Varela, J.}, \au{Shimizu, A.}, \au{Spong, D.~A.}, \au{Garcia, L.} \& \au{Ghai, Y.}} \yr{2021}  \at{{Study of the Alfv\'{e}n eigenmodes stability in CFQS plasma using a Landau closure model}}.  \jt{Nuclear Fusion}  \bvol{61}~(2),  \pg{026023}.

\bibitem[Wei{\ss}e {\em et~al.\/}(2006)Wei{\ss}e, Wellein, Alvermann \& Fehske]{2006Weisse}
{\sc \au{Wei{\ss}e, A.}, \au{Wellein, G.}, \au{Alvermann, A.} \& \au{Fehske, H.}} \yr{2006}  \at{The kernel polynomial method}.  \jt{Reviews of Modern Physics}  \bvol{78}~(1),  \pg{275--306}.

\bibitem[Yakovenko {\em et~al.\/}(2007)Yakovenko, Weller, Werner, Zegenhagen, Fesenyuk \& Kolesnichenko]{2007Yakovenko}
{\sc \au{Yakovenko, Y.~V.}, \au{Weller, A.}, \au{Werner, A.}, \au{Zegenhagen, S.}, \au{Fesenyuk, O.~P.} \& \au{Kolesnichenko, Y.~I.}} \yr{2007}  \at{{Poloidal trapping of the high-frequency Alfv{\'e}n continuum and eigenmodes in stellarators}}.  \jt{Plasma Physics and Controlled Fusion}  \bvol{49}~(4),  \pg{535}.

\bibitem[Zarnstorff {\em et~al.\/}(2001)Zarnstorff, Berry, Brooks, Fredrickson, Fu, Hirshman, Hudson, Ku, Lazarus, Mikkelsen {\em et~al.\/}]{2001Zarnstorff}
{\sc \au{Zarnstorff, M.~C.}, \au{Berry, L.~A.}, \au{Brooks, A.}, \au{Fredrickson, E.}, \au{Fu, G.~Y.}, \au{Hirshman, S.}, \au{Hudson, S.}, \au{Ku, L.~P.}, \au{Lazarus, E.}, \au{Mikkelsen, D.} \& \au{others}} \yr{2001}  \at{{Physics of the compact advanced stellarator NCSX}}.  \jt{Plasma Physics and Controlled Fusion}  \bvol{43}~(12A),  \pg{A237}.

\bibitem[Zhu {\em et~al.\/}(2025)Zhu, Lin \& Bhattacharjee]{2025Zhu}
{\sc \au{Zhu, H.}, \au{Lin, Z.} \& \au{Bhattacharjee, A.}} \yr{2025}  \at{Collisionless zonal-flow dynamics in quasisymmetric stellarators}.  \jt{Journal of Plasma Physics}  \bvol{91}~(1),  \pg{E28}.

\end{thebibliography}

\end{document}